# The Agda Universal Algebra Library
# Part 1: Foundation

**Equality, extensionality, truncation, and dependent types for relations and algebras**

William DeMeo ✉ 🏠 ⓘ
Department of Algebra, Charles University in Prague

## Abstract

The Agda Universal Algebra Library (UALib) is a library of types and programs (theorems and proofs) we developed to formalize the foundations of universal algebra in dependent type theory using the Agda programming language and proof assistant. The UALib includes a substantial collection of definitions, theorems, and proofs from general algebra and equational logic, including many examples that exhibit the power of inductive and dependent types for representing and reasoning about relations, algebraic structures, and equational theories. In this paper we discuss the logical foundations on which the library is built, and describe the types defined in the first 13 modules of the library. Special attention is given to aspects of the library that seem most interesting or challenging from a type theory or mathematical foundations perspective.

## Acknowledgements

The author thanks Cliff Bergman, Hyeyoung Shin, and Siva Somayyajula for supporting and contributing to this project, Adreas Abel for helpful corrections, and three anonymous referees for invaluable feedback on an early draft of this work. Thanks are also owed to Martín Escardó for creating the Type Topology library and teaching us about it at the 2019 Midlands Graduate School in Computing Science [10]. Finally, the AgdaUALib would not exist in its current form without the Agda language, for which we thank the Agda Team (Andreas Abel, Guillaume Allais, Liang-Ting Chen, Jesper Cockx, Nils Anders Danielsson, Víctor López Juan, Ulf Norell, Andrés Sicard-Ramírez, Andrea Vezzosi, and Tesla Ice Zhang).

**Contents**

## 1 Introduction

To support formalization in type theory of research level mathematics in universal algebra and related fields, we present the Agda Universal Algebra Library (AgdaUALib), a software library containing formal statements and proofs of the core definitions and results of universal algebra. The UALib is written in Agda [17], a programming language and proof assistant based on Martin-Löf Type Theory (MLTT) that supports dependent and inductive types.

### 1.1 Motivation

The seminal idea for the AgdaUALib project was the observation that, on the one hand, a number of fundamental constructions in universal algebra can be defined recursively, and theorems about them proved by structural induction, while, on the other hand, inductive and dependent types make possible very precise formal representations of recursively defined objects, which often admit elegant constructive proofs of properties of such objects. An important feature of such proofs in type theory is that they are total functional programs and, as such, they are computable, composable, and machine-verifiable.

Finally, our own research experience has taught us that a proof assistant and programming language (like Agda), when equipped with specialized libraries and domain-specific tactics to automate the proof idioms of our field, can be an extremely powerful and effective asset. As such we believe that proof assistants and their supporting libraries will eventually become indispensable tools in the working mathematician's toolkit.

## 1.2 Attributions and Contributions

The mathematical results described in this paper have well known *informal* proofs. Our main contribution is the formalization, mechanization, and verification of the statements and proofs of these results in dependent type theory using Agda.

Unless explicitly stated otherwise, the Agda source code described in this paper is due to the author, with the following caveat: the UALib depends on the Type Topology library of Martín Escardó [10]. For convenience, we refer to Escardó's library as TypeTopo throughout the paper. For the sake of completeness and clarity, and to keep the paper mostly self-contained, we repeat some definitions from TypeTopo, but in each instance we cite the original source.[2]

## 1.3 Prior art

There have been a number of efforts to formalize parts of universal algebra in type theory prior to ours, most notably

- Capretta [4] (1999) formalized the basics of universal algebra in the Calculus of Inductive Constructions using the Coq proof assistant;
- Spitters and van der Weegen [19] (2011) formalized the basics of universal algebra and some classical algebraic structures, also in the Calculus of Inductive Constructions using the Coq proof assistant, promoting the use of type classes;
- Gunther, et al [11] (2018) developed what seems to be (prior to the UALib) the most extensive library of formal universal algebra to date; in particular, this work includes a formalization of some basic equational logic; also (unlike the UALib) it handles *multisorted* algebraic structures; (like the UALib) it is based on dependent type theory and the Agda proof assistant.

Some other projects aimed at formalizing mathematics generally, and algebra in particular, have developed into very extensive libraries that include definitions, theorems, and proofs about algebraic structures, such as groups, rings, modules, etc. However, the goals of these efforts seem to be the formalization of special classical algebraic structures, as opposed to the general theory of (universal) algebras. Moreover, the part of universal algebra and equational logic formalized in the UALib extends beyond the scope of prior efforts. In particular, the library now includes a proof of Birkhoff's variety theorem. Most other proofs of this theorem that we know of are informal and nonconstructive.[3]

## 1.4 Organization of the paper

In this paper we limit ourselves to the presentation of the core foundational modules of the UALib so that we have space to discuss some of the more interesting type theoretic and foundational issues that arose when developing the library and attempting to represent advanced mathematical notions in type theory and formalize them in Agda. This is the first in a series of three papers describing the AgdaUALib. The second paper ([8]) covers *homomorphisms*, *terms*, and *subalgebras*. The third paper ([9]) covers *free algebras*, *equational classes* of algebras (i.e., *varieties*), and *Birkhoff's HSP theorem*.

---

[2] In the UALib, such instances occur only inside hidden modules that are never actually used, followed immediately by a statement that imports the code in question from its original source.

[3] After completing the formal proof in Agda, we learned about a constructive version of Birkhoff's theorem proved by Carlström in [5]. The latter is presented in the informal style of standard mathematical writing, and as far as we know it was never formalized in type theory and type-checked with a proof assistant. Nonetheless, a comparison of Carlström's proof and the UALib proof would be interesting.

This present paper is organized into three parts as follows. The first part is §2 which introduces the basic concepts of type theory with special emphasis on the way such concepts are formalized in Agda. Specifically, §2.1 introduces *Sigma types* and Agda's hierarchy of *universes*. The important topics of *equality* and *function extensionality* are discussed in §2.2 and §2.3; §2.4 covers inverses and inverse images of functions. In §2.5 we describe a technical problem that one frequently encounters when working in a *noncumulative universe hierarchy* and offer some tools for resolving the type-checking errors that arise from this.

The second part is §3 which covers *relation types* and *quotient types*. Specifically, §3.1 defines types that represent *unary* and *binary* relations as well as *function kernels*. These "discrete relation types," are all very standard. In §3.2 we introduce the (less standard) types that we use to represent *general* and *dependent relations*. We call these "continuous relations" because they can have arbitrary arity (general relations) and they can be defined over arbitrary families of types (dependent relations). In §3.3 we cover standard types for equivalence relations and quotients, and in §3.4 we discuss a family of concepts that are vital to the mechanization of mathematics using type theory; these are the closely related concepts of *truncation*, *sets*, *propositions*, and *proposition extensionality*.

The third part of the paper is §4 which covers the basic domain-specific types offered by the UALib. It is here that we finally get to see some types representing algebraic structures. Specifically, we describe types for *operations* and *signatures* (§4.1), *general algebras* (§4.2), and *product algebras* (§4.3), including types for representing *products over arbitrary classes of algebraic structures*. Finally, we define types for congruence relations and quotient algebras in §4.4.

## 1.5 Resources

We conclude this introduction with some pointers to helpful reference materials. For the required background in Universal Algebra, we recommend the textbook by Clifford Bergman [1]. For the type theory background, we recommend the HoTT Book [18] and Escardó's Introduction to Univalent Foundations of Mathematics with Agda [10].

The following are informed the development of the UALib and are highly recommended.
- *Introduction to Univalent Foundations of Mathematics with Agda*, Escardó [10].
- *Dependent Types at Work*, Bove and Dybjer [2].
- *Dependently Typed Programming in Agda*, Norell and Chapman [16].
- *Formalization of Universal Algebra in Agda*, Gunther, Gadea, Pagano [11].
- *Programming Languages Foundations in Agda*, Philip Wadler [24].

More information about AgdaUALib can be obtained from the following official sources.
- ualib.org (the web site) documents every line of code in the library.
- gitlab.com/ualib/ualib.gitlab.io (the source code) AgdaUALib is open source.[4]
- *The Agda UALib, Part 2: homomorphisms, terms, and subalgebras* [8].
- *The Agda UALib, Part 3: free algebras, equational classes, and Birkhoff's theorem* [9].

---

[4] License: Creative Commons Attribution-ShareAlike 4.0 International License.

The first item links to the official UALib html documentation which includes complete proofs of every theorem we mention here, and much more, including the Agda modules covered in the first and third installments of this series of papers on the UALib.

Finally, readers will get much more out of reading the paper if they download the AgdaUALib from https://gitlab.com/ualib/ualib.gitlab.io, install the library, and try it out for themselves.

## 2 Overture

### 2.1 Preliminaries

This section presents the Overture.Preliminaries module of the AgdaUALib, slightly abridged.[5] Here we define or import the basic types of *Martin-Löf type theory* (MLTT). Although this is standard stuff, we take this opportunity to highlight aspects of the UALib syntax that may differ from that of "standard Agda."

**Logical foundations**

The AgdaUALib is based on a type theory that is the same or very close to the one on which on which Martín Escardó's Type Topology (TypeTopo) Agda library is based. We don't discuss MLTT in great detail here because there are already nice and freely available resources covering the theory. (See, for example, the section A spartan Martin-Löf type theory of the lecture notes by Escardó [10], the ncatlab entry on Martin-Löf dependent type theory, or the HoTT Book [18].)

The objects and assumptions that form the foundation of MLTT are few. There are the *primitive types* ($\mathbb{0}$, $\mathbb{1}$, and $\mathbb{N}$, denoting the empty type, one-element type, and natural numbers), the *type formers* (+, Π, Σ, Id, denoting *binary sum*, *product*, *sum*, and the *identity* type). Each of these type formers is defined by a *type forming rule* which specifies how that type is constructed. Lastly, we have an infinite collection of *type universes* (types of types) and *universe variables* to denote them. Following Escardó, we denote universes in the UALib by upper-case calligraphic letters from the second half of the English alphabet; to be precise, these are $\mathcal{O}$, $\mathcal{Q}$, $\mathcal{R}$, ..., $\mathcal{X}$, $\mathcal{Y}$, $\mathcal{Z}$.[6]

That's all. There are no further axioms or logical deduction (proof derivation) rules needed for the foundation of MLTT that we take as the starting point of the AgdaUALib. The logical semantics come from the propositions-as-types correspondence [15]: propositions and predicates are represented by types and the inhabitants of these types are the proofs of the propositions and predicates. As such, proofs are constructed using the type forming rules. In other words, the type forming rules *are* the proof derivation rules.

To this foundation, we add certain *extensionality principles* when and were we need them. These will be developed as we progress. However, classical axioms such as the *Axiom of Choice* or the *Law of the Excluded Middle* are not needed and are not assumed anywhere in the library. In that sense, all theorems and proofs in the UALib are *constructive* (as defined, e.g., in [13]).

A few specific instances (e.g., the proof of the Noether isomorphism theorems and Birkhoff's HSP theorem) require certain *truncation* assumptions. In such cases, the theory is not *predicative* (as defined, e.g., in [14]). These instances are always clearly identified.

---

[5] For unabridged docs (source code) see https://ualib.gitlab.io/Overture.Preliminaries.html (https://gitlab.com/ualib/ualib.gitlab.io/-/blob/master/UALib/Overture/Preliminaries.lagda).

[6] We avoid using $\mathcal{P}$ as a universe variable because it is used to denote the powerset type in TypeTopo.

**Specifying logical foundations in Agda**

An Agda program typically begins by setting some options and by importing types from existing Agda libraries. Options are specified with the OPTIONS *pragma* and control the way Agda behaves, for example, by specifying the logical axioms and deduction rules we wish to assume when the program is type-checked to verify its correctness. Every Agda program in the UALib begins with the following line.

$$\{\text{-\# OPTIONS --without-K --exact-split --safe \#-}\} \tag{1}$$

These options control certain foundational assumptions that Agda makes when type-checking the program to verify its correctness.

1. *–without-K* disables Streicher's K axiom, which makes Agda compatible with *proof-relevant* type theories; see the discussion of proof-relevance in § 3.4; see also [20, 7];
2. *–exact-split* makes Agda accept only definitions that are *judgmental* equalities; see [22];
3. *–safe* ensures that nothing is postulated outright—every non-MLTT axiom has to be an explicit assumption (e.g., an argument to a function or module); see [21, 22].

Throughout this paper we take assumptions 1–3 for granted without mentioning them explicitly.

**Agda Modules**

The OPTIONS pragma is usually followed by the start of a module. For example, the Overture.Preliminaries module begins with the following line.

    module Overture.Preliminaries where

Sometimes we want to declare parameters that will be assumed throughout the module. For instance, when working with algebras, we often assume they come from a particular fixed signature, and this signature is something we could fix as a parameter at the start of a module. Thus, we might start an *anonymous submodule* of the main module with a line like[7]

    module _ {$S$ : Signature $\mathcal{O}$ $\mathcal{V}$} where

Such a module is called *anonymous* because an underscore appears in place of a module name. Agda determines where a submodule ends by indentation. This can take some getting used to, but after a short time it will feel very natural. The main module of a file must have the same name as the file, without the `.agda` or `.lagda` file extension. The code inside the main module is not indented. Submodules are declared inside the main module and code inside these submodules must be indented to a fixed column. As long as the code is indented, Agda considers it part of the submodule. A submodule is exited as soon as a nonindented line of code appears.

**Universes in Agda**

For the very small amount of background we require about the notion of *type universe* (or *level*), we refer the reader to the brief section on universe-levels in the Agda documentation.[8]

    Throughout the AgdaUALib we use many of the nice tools that Martín Escardó has developed and made available in TypeTopo library of Agda code for the *Univalent Foundations* of math-

---

[7] The Signature type will be defined in Section 4.1.
[8] See https://agda.readthedocs.io/en/v2.6.1.3/language/universe-levels.html.

ematics.[9] The first of these is the Universes module which we import as follows.

    open import Universes public

Since we use the public directive, the Universes module will be available to all modules that import the present module (Overture.Preliminaries). This module declares symbols used to denote universes. As mentioned, we adopt Escardó's convention of denoting universes by capital calligraphic letters, and most of the ones we use are already declared in Universes; those that are not are declared as follows.

    variable $\mathcal{O}$ $\mathcal{X}$ $\mathcal{Y}$ $\mathcal{Z}$ : Universe

The Universes module also provides alternative syntax for the primitive operations on universes that Agda supports. Specifically, the ̇ operator maps a universe level $\mathcal{U}$ to the type Set $\mathcal{U}$, and the latter has type Set (lsuc $\mathcal{U}$). The Agda level lzero is renamed $\mathcal{U}_0$, so $\mathcal{U}_0$ ̇ is an alias for Set lzero. Thus, $\mathcal{U}$ ̇ is simply an alias for Set $\mathcal{U}$, and we have Set $\mathcal{U}$ : Set (lsuc $\mathcal{U}$). Finally, Set (lsuc lzero) is equivalent to Set $\mathcal{U}_0^+$, which we (and Escardó) denote by $\mathcal{U}_0^+$ ̇.

To justify the introduction of this somewhat nonstandard notation for universe levels, Escardó points out that the Agda library uses Level for universes (so what we write as $\mathcal{U}$ ̇ is written Set $\mathcal{U}$ in standard Agda), but in univalent mathematics the types in $\mathcal{U}$ ̇ need not be sets, so the standard Agda notation can be a bit confusing, especially to newcomers.

There will be many occasions calling for a type living in a universe at the level that is the least upper bound of two universe levels, say, $\mathcal{U}$ and $\mathcal{V}$. The universe level $\mathcal{U} \sqcup \mathcal{V}$ denotes this least upper bound. Here ⊔ is an Agda primitive designed for precisely this purpose.

## Dependent types

### Sigma types (dependent pairs)

Given universes $\mathcal{U}$ and $\mathcal{V}$, a type $A : \mathcal{U}$ ̇, and a type family $B : A \to \mathcal{V}$ ̇, the *Sigma type* (or *dependent pair type*, or *dependent product type*) is denoted by $\Sigma\ x : A\ ,\ B\ x$ and generalizes the Cartesian product $A \times B$ by allowing the type $B\ x$ of the second argument of the ordered pair $(x , y)$ to depend on the value $x$ of the first. That is, an inhabitant of the type $\Sigma\ x : A\ ,\ B\ x$ is a pair $(x , y)$ such that $x : A$ and $y : B\ x$.

The dependent product type is defined in TypeTopo in a standard way. For pedagogical purposes we repeat the definition here.[10]

    record $\Sigma$ {$\mathcal{U}$ $\mathcal{V}$} {$A : \mathcal{U}$ ̇} ($B : A \to \mathcal{V}$ ̇) : $\mathcal{U} \sqcup \mathcal{V}$ ̇ where
      constructor _,_
      field
        pr$_1$ : $A$
        pr$_2$ : $B$ pr$_1$

Agda's default syntax for this type is $\Sigma\ \lambda(x : A) \to B$, but we prefer the notation $\Sigma\ x : A\ ,\ B$,

---

[9] Escardó has written an outstanding set of notes called Introduction to Univalent Foundations of Mathematics with Agda, which we highly recommend to anyone looking for more details than we provide here about MLTT and Univalent Foundations/HoTT in Agda. [10].

[10] In the UALib we put such redundant definitions inside "hidden" modules so that they doesn't conflict with the original definitions which we import and use. It may seem odd to define something in a hidden module only to import and use an alternative definition, but we do this in order to exhibit all of the types on which the UALib depends while ensuring that this cannot be misinterpreted as a claim to originality.

which is closer to the syntax in the preceding paragraph, and will be familiar to readers of the HoTT book [18], for example. Fortunately, TypeTopo makes the preferred notation available with the following type definition and syntax declaration (see [10, Σ types]).[11]

```
-Σ : {𝒰 𝒱 : Universe} (A : 𝒰 ˙ ) (B : A → 𝒱 ˙ ) → 𝒰 ⊔ 𝒱 ˙
-Σ A B = Σ B

syntax -Σ A (λ x → B) = Σ x : A , B
```

A special case of the Sigma type is the one in which the type $B$ doesn't depend on $A$. This is the usual Cartesian product, defined in Agda as follows.

```
_×_ : 𝒰 ˙ → 𝒱 ˙ → 𝒰 ⊔ 𝒱 ˙
A × B = Σ x : A , B
```

### Pi types (dependent functions)

Given universes $\mathcal{U}$ and $\mathcal{V}$, a type $A : \mathcal{U}\ ˙$, and a type family $B : A \to \mathcal{V}\ ˙$, the *Pi type* (or *dependent function type*) is denoted by $\Pi\ x : A\ ,\ B\ x$ and generalizes the function type $A \to B$ by letting the type $B\ x$ of the codomain depend on the value $x$ of the domain type. The dependent function type is defined in TypeTopo in a standard way. For the reader's benefit, however, we repeat the definition here.

```
Π : {A : 𝒰 ˙ } (A : A → 𝒲 ˙ ) → 𝒰 ⊔ 𝒲 ˙
Π {A} A = (x : A) → A x
```

To make the syntax for Π conform to the standard notation for Pi types, Escardó uses the same trick as the one used above for Sigma types.[11]

```
-Π : (A : 𝒰 ˙ )(B : A → 𝒲 ˙ ) → 𝒰 ⊔ 𝒲 ˙
-Π A B = Π B

syntax -Π A (λ x → b) = Π x : A , b
```

Once we have studied the types (defined in TypeTopo and repeated here for convenience and illustration purposes), the original definitions are imported like so.

```
open import Sigma-Type public
open import MGS-MLTT using (pr₁; pr₂; _×_; -Σ; Π; -Π) public
```

### Projection notation

The definition of Σ (and thus ×) includes the fields $pr_1$ and $pr_2$ representing the first and second projections out of the product. Sometimes we prefer to denote these projections by $|\_|$ and $\|\_\|$, respectively. However, for emphasis or readability we alternate between these and the following standard notations: $pr_1$ and fst for the first projection, $pr_2$ and snd for the second. We define these alternative notations for projections as follows.

```
module _ {𝒰 : Universe}{A : 𝒰 ˙ }{B : A → 𝒱 ˙} where

 |_| fst : Σ B → A
```

---

[11] **Attention!** The symbol : that appears in the special syntax defined here for the Σ type, and below for the Π type, is not the ordinary colon; rather, it is the symbol obtained by typing \:4 in agda2-mode.

```
             | x , y | = x
             fst (x , y) = x

             ∥_∥ snd : (z : Σ B) → B (pr₁ z)
             ∥ x , y ∥ = y
             snd (x , y) = y
```

**Remarks**.
- We place these definitions (of |_|, fst, ∥_∥ and snd) inside an *anonymous module*, which is a module that begins with the module keyword followed by an underscore character (instead of a module name). The purpose is to move some of the postulated typing judgments—the "parameters" of the module (e.g., 𝒰 : Universe)—out of the way so they don't obfuscate the definitions inside the module. In library documentation, such as the present paper, we often omit such module directives. In contrast, the collection of html pages at ualib.org, which is the most current and comprehensive documentation of the UALib, omits nothing.
- As the four definitions above make clear, multiple inhabitants of a single type (e.g., |_| and fst) may be declared on the same line.

## 2.2  Equality

This section presents the Overture.Equality module of the AgdaUALib, slightly abridged.[12]

### Definitional equality

Here we discuss the basic but important type of MLTT called *definitional equality*. This concept is most understood, at least heuristically, with the following slogan: "Definitional equality is the substitution-preserving equivalence relation generated by definitions." We will make this precise below, but first let us quote from a primary source. Per Martin-Löf offers the following definition in [12, §1.11] (italics added):[13]

> *Definitional equality* is defined to be the equivalence relation, that is, reflexive, symmetric and transitive relation, which is generated by the principles that a definiendum is always definitionally equal to its definiens and that definitional equality is preserved under substitution.

To be sure we understand what this means, let := denote the relation with respect to which $x$ is related to $y$ (denoted $x := y$) if and only if *$y$ is the definition of $x$*. Then the definitional equality relation ≡ is the reflexive, symmetric, transitive, substitutive closure of :=. By *subsitutive closure* we mean closure under the following *substitution rule*.

$$\frac{\{A : \mathcal{U} \;\cdot\}\; \{B : A \to \mathcal{W} \;\cdot\}\; \{x\, y : A\} \quad x \equiv y}{B\, x \equiv B\, y} \; (subst)$$

The datatype used in the UALib to represent definitional equality is imported from the Identity-Type module of TypeTopo, but apart from superficial syntactic differences, it is equivalent to the standard *Paulin-Mohring style identity type* found in most other Agda libraries. We repeat the definition here for easy reference.

---

[12] For unabridged docs (source code) see https://ualib.gitlab.io/Overture.Equality.html (https://gitlab.com/ualib/ualib.gitlab.io/-/blob/master/UALib/Overture/Equality.lagda).

[13] The *definiendum* is the left-hand side of a defining equation, the *definiens* is the right-hand side. For readers who have never generated an equivalence relation: the *reflexive closure* of $R \subseteq A \times A$ is the union of $R$ and all pairs of the form $(a , a)$; the *symmetric closure* is the union of $R$ and its inverse $\{(y , x) : (x , y) \in R\}$; we leave it to the reader to come up with the correct definition of transitive closure.

> data $\_\equiv\_$ $\{\mathcal{U}\}$ $\{A : \mathcal{U}\ \cdot\ \}$ : $A \to A \to \mathcal{U}\ \cdot$ where refl : $\{x : A\} \to x \equiv x$

Whenever we need to complete a proof by simply asserting that $x$ is *definitionally equal* to itself, we invoke refl. If we need to make explicit the implicit argument $x$, then we use refl $\{x = x\}$.

**Assumed module contexts**

Before proceeding, a word about a special convention we adopt in the sequel is in order. To reduce reader strain, we often omit easily inferred typing judgments which would normally appear in the list of parameters of a module or at the start of a type definition, though we sometimes make an announcement like the following (which applies to the present section):

> *Unless otherwise indicated, the prevailing context in this section is given by*
>
> module $\_$ $\{\mathcal{U} : $ Universe$\}$ $\{A : \mathcal{U}\ \cdot\}$ where which means that all code in the current section is to be interpreted as occurring inside such an anonymous module, where the given typing judgments are taken for granted.

**Identity is an equivalence relation**

The relation $\equiv$ just defined is naturally an equivalence relation, and the formal proof of this fact is trivial. Indeed, we don't need to prove reflexivity, since that is the defining property of $\equiv$, and the proofs of symmetry and transitivity are immediate.

> $\equiv$-sym : $\{x\ y : A\} \to x \equiv y \to y \equiv x$
> $\equiv$-sym refl $=$ refl
>
> $\equiv$-trans : $\{x\ y\ z : A\} \to x \equiv y \to y \equiv z \to x \equiv z$
> $\equiv$-trans refl refl $=$ refl

We prove that $\equiv$ obeys the substitution rule (subst) in the next subsection (see ap), but first we define some syntactic sugar that will make it easier to apply symmetry and transitivity of $\equiv$ in proofs.[14]

> $\_^{-1}$ : $\{x\ y : A\} \to x \equiv y \to y \equiv x$
> $p\ ^{-1} = \equiv$-sym $p$

If we have a proof $p : x \equiv y$, and we need a proof of $y \equiv x$, then instead of $\equiv$-sym $p$ we can use the more intuitive $p\ ^{-1}$. Similarly, the following syntactic sugar makes abundant appeals to transitivity easier to stomach.

> $\_\cdot\_$ : $\{x\ y\ z : A\} \to x \equiv y \to y \equiv z \to x \equiv z$
> $p \cdot q = \equiv$-trans $p\ q$

**Transport (substitution)**

Alonzo Church characterized equality by declaring two things equal if and only if no property (predicate) can distinguish them (see [6]). In other terms, $x$ and $y$ are equal if and only if for all $P$ we have $P\ x \to P\ y$. One direction of this implication is sometimes called *substitution* or *transport* or *transport along an identity*. It asserts the following: if two objects are equal and one of them satisfies a given predicate, then so does the other. A type representing this notion is defined, along with the (polymorphic) identity function, in the MGS-MLTT module of TypeTopo as follows.[15]

---

[14] **Unicode Hints** (agda2-mode): \^-\^1 ⇝ $^{-1}$; \Mii\Mid ⇝ *id*; \. ⇝ ·. In general, for information about a character, place the cursor on the character and type M-x describe-char (or M-x h d c).

```
id : {𝒰 : Universe} (A : 𝒰 ˙ ) → A → A
id A = λ x → x

transport : {A : 𝒰 ˙ } (B : A → 𝒲 ˙ ) {x y : A} → x ≡ y → B x → B y
transport B (refl {x = x}) = id (B x)
```

A function is well-defined if and only if it maps equivalent elements to a single element and we often use this nature of functions in Agda proofs. It is equivalent to the substitution rule (subst) we defined in the last section. If we have a function $f\colon A \to B$, two elements $x\ y : A$ of the domain, and an identity proof $p : x \equiv y$, then we obtain a proof of $f\ x \equiv f\ y$ by simply applying the ap function like so, ap $f\ p : f\ x \equiv f\ y$. Escardó defines ap in TypeTopo as follows.

```
ap : {A : 𝒰 ˙ }{B : 𝒱 ˙ } (f : A → B){a b : A} → a ≡ b → f a ≡ f b
ap f {a} p = transport (λ - → f a ≡ f -) p (refl {x = f a})
```

This establishes that our definitional equality satisfies the substitution rule (subst).

Here's a useful variation of ap that we borrow from the `Relation/Binary/Core.agda` module of the Agda Standard Library (transcribed into TypeTopo/UALib).

```
cong-app : {A : 𝒰 ˙ }{B : A → 𝒲 ˙ }{f g : Π B} → f ≡ g → (a : A) → f a ≡ g a
cong-app refl _ = refl
```

### 2.3 Function extensionality

This section presents the Overture.Extensionality module of the AgdaUALib, slightly abridged.[16] This brief introduction to *function extensionality* is intended for novices. Those already familiar with the concept might wish to skip to the next subsection.

What does it mean to say that two functions $f\ g\colon X \to Y$ are equal? Suppose $f$ and $g$ are defined on $X = \mathbb{Z}$ (the integers) as follows: $fx := x + 2$ and $gx := ((2 * x) - 8)/2 + 6$. Should we call $f$ and $g$ equal? Are they the "same" function? What does that even mean?

It's hard to resist the urge to reduce $g$ to $x + 2$ and proclaim that $f$ and $g$ are equal. Indeed, this is often an acceptable answer and the discussion normally ends there. In the science of computing, however, more attention is paid to equality, and with good reason.

We can probably all agree that the functions $f$ and $g$ above, while not syntactically equal, do produce the same output when given the same input so it seems fine to think of the functions as the same, for all intents and purposes. But we should ask ourselves at what point do we notice or care about the difference in the way functions are defined? What if we had started out this discussion with two functions $f$ and $g$ both of which take a list as argument and produce as output a correctly sorted version of the input list? Suppose $f$ is defined using the merge sort algorithm, while $g$ uses quick sort. Probably few of us would call $f$ and $g$ the "same" in this case.

---

[15] Including every line of code of the AgdaUALib in this paper would result in an unbearable reading experience. We include all significant sections of code from the first 13 modules, but we omit lines indicating that redundant definitions of functions (e.g., transport and ap) occur inside named "hidden" modules. We also omit lines importing the original definitions of such duplicate definitions from TypeTopo library.

[16] For unabridged docs (source code) see https://ualib.gitlab.io/Overture.Extensionality.html (https://gitlab.com/ualib/ualib.gitlab.io/-/blob/master/UALib/Overture/Extensionality.lagda).

In the examples above, it is common to say that the two functions are *extensionally equal*, since they produce the same *external* output when given the same input, but they are not *intensionally equal*, since their *internal* definitions differ. In the next subsection we describe types that manifest this idea of *extensional equality of functions*, or *function extensionality*.[17]

### Types for postulating function extensionality

As explained above, a natural notion of function equality is defined as follows: $f$ and $g$ are said to be *pointwise equal* provided $\forall x \to fx \equiv gx$. Here is how this is expressed in type theory (e.g., in TypeTopo).

```
_∼_ : {A : 𝒰 ˙} {B : A → 𝒲 ˙} → Π B → Π B → 𝒰 ⊔ 𝒲 ˙
f ∼ g = ∀ x → f x ≡ g x
```

*Function extensionality* is the principle that pointwise equal functions are *definitionally equal*; that is, $\forall f\ g\ (f \sim g \to f \equiv g)$. In type theory this principle is represented by the types funext (for nondependent functions) and dfunext (for dependent functions) ([18, §2.9]). For example, the latter is defined as follows.[18]

```
dfunext : ∀ 𝒰 𝒲 → (𝒰 ⊔ 𝒲) ⁺ ˙
dfunext 𝒰 𝒲 = {A : 𝒰 ˙}{B : A → 𝒲 ˙}{f g : ∀(x : A) → B x} → f ∼ g → f ≡ g
```

In informal settings, this so-called *point-wise equality of functions* is typically what one means when one asserts that two functions are "equal."[19] However, it is important to keep in mind the following fact: *function extensionality is known to be neither provable nor disprovable in Martin-Löf type theory. It is an independent statement.*. [10]

The next type defines a converse of function extensionality for dependent function types (cf. cong-app in Overture.Equality and [18, (2.9.2)]).

```
happly : {A : 𝒰 ˙}{B : A → 𝒲 ˙}(f g : Π B) → f ≡ g → f ∼ g
happly _ _ refl _ = refl
```

Though it may seem obvious to some readers, we wish to emphasize the important conceptual distinction between two different forms of type definitions by comparing the definitions of dfunext and happly. In the definition of dfunext, the codomain is a parameterized type, namely, $\mathcal{U}^+ \sqcup \mathcal{V}^+$ ˙, and the right-hand side of the defining equation of dfunext is an assertion (which may or may not hold). In the definition of happly, the codomain is an assertion, namely, $f \sim g$, and the right-hand side of the defining equation is a proof of this assertion. As such, happly is a *proof object*; it *proves* (or *inhabits the type that represents*) the proposition asserting that definitionally equivalent functions are pointwise equal. In contrast, dfunext is a type, and we may or may not wish to postulate an inhabitant of this type. That is, we could postulate that function extensionality holds by assuming we have a witness, say, $fe$ : dfunext $\mathcal{U}$ $\mathcal{V}$, but as noted above the existence of such a witness cannot be proved in MLTT.

---

[17] Most of these types are already defined in TypeTopo, so the UALib imports the definitions from there; as usual, we redefine some of these types here for the purpose of explication.

[18] Previous versions of the UALib made heavy use of a *global function extensionality principle* which asserts that function extensionality holds at all universe levels. However, we removed all instances of global function extensionality in favor of local applications of the principle. This makes transparent precisely how and where the library depends on function extensionality. The price we pay for this precision is a proliferation of extensionality postulates. Eventually we will likely be able to remove these postulates with other approach to extensionality; e.g., *univalence* and/or Cubical Agda.

[19] In fact, if one assumes the *univalence axiom* of Homotopy Type Theory [18], then point-wise equality of functions is equivalent to definitional equality of functions. See the section "Function extensionality from univalence" of [10].

Finally, a useful alternative for expressing dependent function extensionality, which is essentially equivalent to dfunext, is to assert that happly is actually an *equivalence* in a sense that we now describe. This requires a few definitions from the MGS-Equivalences module of TypeTopo. First, a type is a *singleton* if it has exactly one inhabitant and a *subsingleton* if it has at most one inhabitant.

```
is-center : (A : 𝒰 ˙) → A → 𝒰 ˙
is-center A c = (x : A) → c ≡ x

is-singleton : 𝒰 ˙ → 𝒰 ˙
is-singleton A = Σ c : A , is-center A c

is-subsingleton : 𝒰 ˙ → 𝒰 ˙
is-subsingleton A = (x y : A) → x ≡ y
```

Next, we consider the type is-equiv which is used to assert that a function is an equivalence in the sense that we now describe. This requires the concept of a *fiber* of a function, which can be represented as a Sigma type whose inhabitants denote inverse images of points in the codomain of the given function.

```
fiber : {A : 𝒰 ˙} {B : 𝒲 ˙} (f : A → B) → B → 𝒰 ⊔ 𝒲 ˙
fiber {A} f y = Σ x : A , f x ≡ y
```

A function is called an *equivalence* if all of its fibers are singletons.

```
is-equiv : {A : 𝒰 ˙} {B : 𝒲 ˙} → (A → B) → 𝒰 ⊔ 𝒲 ˙
is-equiv f = ∀ y → is-singleton (fiber f y)
```

Finally we are ready to fulfill the promise of a type that provides an alternative means of postulating function extensionality.

```
hfunext : ∀ 𝒰 𝒲 → (𝒰 ⊔ 𝒲)⁺ ˙
hfunext 𝒰 𝒲 = {A : 𝒰 ˙}{B : A → 𝒲 ˙} (f g : Π B) → is-equiv (happly f g)
```

## 2.4 Inverses

This section presents the Overture.Inverses module of the AgdaUALib, slightly abridged.[20] Assume the following typing judgments: $\{\mathcal{U}\ \mathcal{W}\ :$ Universe$\}\{A : \mathcal{U}\ ˙\}\{B : \mathcal{W}\ ˙\}$.
We begin by defining an inductive type that represents the *inverse image* of a function.

```
data Image_∋_ (f : A → B) : B → 𝒰 ⊔ 𝒲 ˙ where
  im : (x : A) → Image f ∋ f x
  eq : (b : B) → (a : A) → b ≡ f a → Image f ∋ b
```

Next we verify that the type behaves as we expect.

---

[20] For unabridged docs (source code) see https://ualib.gitlab.io/Overture.Inverses.html
(https://gitlab.com/ualib/ualib.gitlab.io/-/blob/master/UALib/Overture/Inverses.lagda).

```
ImageIsImage : (f : A → B)(b : B)(a : A) → b ≡ f a → Image f ∋ b
ImageIsImage f b a b≡fa = eq b a b≡fa
```

An inhabitant of Image $f \ni b$ is a pair $(a , p)$, where $a : A$, and $p$ is a proof that $f$ maps $a$ to $b$; that is, $p : b \equiv f\, a$. Since the proof that $b$ belongs to the image of $f$ is always accompanied by a witness $a : A$, we can actually *compute* a pseudoinverse of $f$. This will be a function that takes an arbitrary $b : B$ and a (*witness*, *proof*)-pair, $(a , p) :$ Image $f \ni b$, and returns $a$.

```
Inv : (f : A → B){b : B} → Image f ∋ b → A
Inv f {.(f a)} (im a) = a
Inv f (eq _ a _) = a
```

We can prove that Inv $f$ is the *right-inverse* of $f$, as follows.

```
InvIsInv : (f : A → B){b : B}(q : Image f ∋ b) → f(Inv f q) ≡ b
InvIsInv f {.(f a)} (im a) = refl
InvIsInv f (eq _ _ p) = p ⁻¹
```

### Epics (surjective functions)

We represent an *epic* (or *surjective*) function from $A$ to $B$ as an inhabitant of the following type.

```
Epic : (A → B) → 𝒰 ⊔ 𝒲 ̇
Epic f = ∀ y → Image f ∋ y
```

With the next definition, we can represent the *right-inverse* of an epic function.

```
EpicInv : (f : A → B) → Epic f → B → A
EpicInv f fE b = Inv f (fE b)
```

The right-inverse of $f$ is obtained by applying EpicInv to $f$ along with a proof of Epic $f$. To see that this does indeed give the right-inverse we prove the EpicInvIsRightInv lemma below. This requires function composition, denoted ∘ and defined in TypeTopo library.

```
_∘_ : {C : B → 𝒲 ̇ } → Π C → (f : A → B) → (x : A) → C (f x)
g ∘ f = λ x → g (f x)
```

Note that the next proof requires function extensionality, which we postulate a module declaration like this: module _ {𝒰 𝒲 : Universe}{fe : funext 𝒲 𝒲}{A : 𝒰 ̇}{B : 𝒲 ̇} where

```
EpicInvIsRightInv : (f : A → B)(fE : Epic f) → f ∘ (EpicInv f fE) ≡ id B
EpicInvIsRightInv f fE = fe (λ x → InvIsInv f (fE x))
```

### Monics (injective functions)

We say that a function $g : A \to B$ is *monic* (or *injective*) if it does not map distinct elements to a common point. The following types manifest this property and prove that monic functions have left-inverses.

```
Monic : (g : A → B) → 𝒰 ⊔ 𝒲 ̇
Monic g = ∀ a₁ a₂ → g a₁ ≡ g a₂ → a₁ ≡ a₂

MonicInv : (f : A → B) → Monic f → (b : B) → Image f ∋ b → A
MonicInv f _ = λ b imfb → Inv f imfb
```

    MonicInvIsLeftInv : $\{f : A \to B\}\{fM : $ Monic $f\}\{x : A\} \to ($MonicInv$\ f\ fM)(f\ x)($im$\ x) \equiv x$
    MonicInvIsLeftInv $=$ refl

### Embeddings

The is-embedding type is defined in TypeTopo in the following way.

    is-embedding : $(A \to B) \to \mathcal{U} \sqcup \mathcal{W}\ \cdot$
    is-embedding $f = \forall\ b \to$ is-subsingleton (fiber $f\ b$)

Thus, is-embedding $f$ asserts that $f$ is a function all of whose fibers are subsingletons. Observe that an embedding is not simply an injective map. However, if we assume that the codomain $B$ has *unique identity proofs* (UIP), then we can prove that a monic function into $B$ is an embedding. We discuss the UIP principle in §3.4 where we present the Relations.Truncation module.

Finding a proof that a function is an embedding isn't always easy, but one path that is often straightforward is to first prove that the function is invertible and then invoke the following theorem.

    invertibles-are-embeddings : $(f : A \to B) \to$ invertible $f \to$ is-embedding $f$
    invertibles-are-embeddings $f\ fi =$ equivs-are-embeddings $f$ (invertibles-are-equivs $f\ fi$)

Finally, embeddings are monic; from a proof $p$ : is-embedding $f$ that $f$ is an embedding we can construct a proof of Monic $f$. We confirm this as follows.

    embedding-is-monic : $(f : A \to B) \to$ is-embedding $f \to$ Monic $f$
    embedding-is-monic $f\ femb\ x\ y\ fxfy =$ ap pr$_1$ $((femb\ (f\ x))$ fx fy$)$ where
      fx fy : fiber $f\ (f\ x)$
      fx $= x$ , refl
      fy $= y$ , $(fxfy\ ^{-1})$

## 2.5 Lifts

This section presents the Overture.Lifts module of the AgdaUALib, slightly abridged.[21]

### Agda's universe hierarchy

The hierarchy of universes in Agda is structured as follows: $\mathcal{U}\ \cdot : \mathcal{U}\ ^+\ \cdot,\ \mathcal{U}\ ^+\ \cdot : \mathcal{U}\ ^{++}\ \cdot$, etc.[22] This means that the universe $\mathcal{U}\ \cdot$ has type $\mathcal{U}\ ^+\ \cdot$, and $\mathcal{U}\ ^+\ \cdot$ has type $\mathcal{U}\ ^{++}\ \cdot$, and so on. It is important to note, however, this does *not* imply that $\mathcal{U}\ \cdot : \mathcal{U}\ ^{++}\ \cdot$. In other words, Agda's universe hierarchy is *noncumulative*. This can be advantageous as it becomes possible to treat universe levels more generally and precisely. On the other hand, a noncumulative hierarchy can sometimes make it seem unduly difficult to convince Agda that a program or proof is correct. To help us overcome this technical issue, there are some general universe lifting and lowering functions, which we describe in the next subsection. In Section 4.2 we will define a couple of domain-specific analogs of these tools. Later, in the modules presented in [8, 9], we prove various properties that make these effective mechanisms for resolving universe level problems when working with algebra types.

---

[21] For unabridged docs (source code) see https://ualib.gitlab.io/Overture.Lifts.html
(https://gitlab.com/ualib/ualib.gitlab.io/-/blob/master/UALib/Overture/Lifts.lagda).

[22] Recall, from the Overture.Preliminaries module (§2.1), the special notation we use to denote Agda's *levels* and *universes*.

**Lifting and lowering**

Let us be more concrete about what is at issue here by considering a typical example. Agda frequently encounters errors during the type-checking process and responds by printing an error message. Often the message has the following form.

```
Birkhoff.lagda:498,20-23
𝒰 != 𝓞 ⊔ 𝒱 ⊔ (𝒰 ⁺) when checking that... has type...
```

This error message means that Agda encountered the universe $\mathcal{U}$ on line 498 (columns 20–23) of the file Birkhoff.lagda, but was expecting to find the universe $\mathcal{O} \sqcup \mathcal{V} \sqcup \mathcal{U}^+$ instead.

The general Lift record type that we now describe makes these situations easier to deal with. It takes a type inhabiting some universe and embeds it into a higher universe and, apart from syntax and notation, it is equivalent to the Lift type one finds in the Level module of the Agda Standard Library.

```
record Lift {𝒲 𝒰 : Universe} (A : 𝒰 ˙) : 𝒰 ⊔ 𝒲 ˙ where
  constructor lift
  field lower : A
open Lift
```

The point of having a ramified hierarchy of universes is to avoid Russell's paradox, and this would be subverted if we were to lower the universe of a type that wasn't previously lifted. However, we can prove that if an application of lower is immediately followed by an application of lift, then the result is the identity transformation. Similarly, lift followed by lower is the identity.

```
lift∼lower : {𝒲 𝒰 : Universe}{A : 𝒰 ˙} → lift ∘ lower ≡ id (Lift{𝒲} A)
lift∼lower = refl

lower∼lift : {𝒲 𝒰 : Universe}{A : 𝒰 ˙} → lower{𝒲}{𝒰} ∘ lift ≡ id A
lower∼lift = refl
```

The proofs are trivial. Nonetheless, we'll come across some holes these types can fill.

## 3  Relation Types

We now present the AgdaUALib's Relations module and its submodules. In §3.1 we define types that represent *unary* and *binary relations*, which we refer to as "discrete relations" to contrast them with the ("continuous") *general* and *dependent relations* that we introduce in §3.2. We call the latter "continuous relations" because they can have arbitrary arity (general relations) and they can be defined over arbitrary families of types (dependent relations).

### 3.1  Discrete relations

This section presents the Relations.Discrete module of the AgdaUALib, slightly abridged.[23]

---

[23] For unabridged docs (source code) see https://ualib.gitlab.io/Relations.Discrete.html (https://gitlab.com/ualib/ualib.gitlab.io/-/blob/master/UALib/Relations/Discrete.lagda).

**Unary relations**

In set theory, given two sets $A$ and $P$, we say that $P$ is a *subset* of $A$, and we write $P \subseteq A$, just in case $\forall\ x\ (x \in P \to x \in A)$. We need a mechanism for representing this notion in Agda. A typical approach is to use a *unary predicate* type which we will denote by Pred and define as follows. Given two universes $\mathcal{U}$ $\mathcal{W}$ and a type $A : \mathcal{U}\ \cdot$, the type Pred $A$ $\mathcal{W}$ represents *properties* that inhabitants of $A$ may or may not satisfy. We write $P$ : Pred $A$ $\mathcal{U}$ to represent the collection of inhabitants of $A$ that satisfy (or belong to) $P$. Here is the definition.[24]

```
Pred : 𝒰 ˙ → (𝒲 : Universe) → 𝒰 ⊔ 𝒲 ⁺ ˙
Pred A 𝒲 = A → 𝒲 ˙
```

This is a general unary predicate type, but by taking the codomain to be Bool = $\{0, 1\}$, we obtain the usual interpretation of set membership; that is, for $P$ : Pred $A$ Bool and for each $x : A$, we would interpret $P\ x \equiv 0$ to mean $x \notin A$, and $P\ x \equiv 1$ to mean $x \in A$.

The UALib includes types that represent the *element inclusion* and *subset inclusion* relations from set theory. For example, given a predicate P, we may represent that "$x$ belongs to P" or that "$x$ has property P," by writing either $x \in P$ or $P\ x$. The definition of $\in$ is standard, as is the definition of $\subseteq$ for the *subset* relation; nonetheless, here they are.[24]

```
_∈_ : A → Pred A 𝒲 → 𝒲 ˙
x ∈ P = P x

_⊆_ : Pred A 𝒲 → Pred A 𝒳 → 𝒰 ⊔ 𝒲 ⊔ 𝒳 ˙
P ⊆ Q = ∀ {x} → x ∈ P → x ∈ Q
```

**Predicates toolbox**

Here is a small collection of tools that will come in handy later. The first is an inductive type that represents *disjoint union*.[25]

```
data _⊎_ (A : 𝒰 ˙) (B : 𝒲 ˙) : 𝒰 ⊔ 𝒲 ˙ where
  inj₁ : (x : A) → A ⊎ B
  inj₂ : (y : B) → A ⊎ B
```

And this can be used to define a type representing *union*, as follows.

```
_∪_ : Pred A 𝒲 → Pred A 𝒳 → Pred A (𝒲 ⊔ 𝒳)
P ∪ Q = λ x → x ∈ P ⊎ x ∈ Q
```

Next we define convenient notation for asserting that the image of a function (the first argument) is contained in a predicate (the second argument).

```
Im_⊆_ : (A → B) → Pred B 𝒳 → 𝒰 ⊔ 𝒳 ˙
Im f ⊆ S = ∀ x → f x ∈ S
```

The *empty set* is naturally represented by the *empty type*, 𝟎, and the latter is defined in the Empty-Type module of TypeTopo.[25,26]

---

[24] cf. `Relation/Unary.agda` in the Agda Standard Library.
[25] **Unicode Hints**. In agda2-mode, `\u+` ⇝ ⊎, `\b0` ⇝ 𝟎, `\B0` ⇝ 𝟬.
[26] The empty type is an inductive type with no constructors; that is, data 𝟎 {𝒰} : 𝒰 ˙ where – *(empty body)*.

∅ : Pred $A$ $\mathcal{U}_0$
∅ _ = 𝟘

We close our little predicates toolbox with a natural way to encode *singletons*.

{ _ } : $A →$ Pred $A$ _
{ $x$ } = $x \equiv$_

### Binary Relations

In set theory, a binary relation on a set $A$ is simply a subset of the Cartesian product $A \times A$. As such, we could model such a relation as a (unary) predicate over the product type $A \times A$, or as an inhabitant of the function type $A \to A \to \mathcal{W}$ ˙ (for some universe $\mathcal{W}$). Note, however, this is not the same as a unary predicate over the function type $A \to A$ since the latter has type $(A \to A) \to \mathcal{W}$ ˙, while a binary relation should have type $A \to (A \to \mathcal{W}$ ˙$)$.

A generalization of the notion of binary relation is a *relation from $A$ to $B$*, which we define first and treat binary relations on a single $A$ as a special case.

REL : $\mathcal{U}$ ˙ $\to \mathcal{W}$ ˙ $\to (\mathcal{X} :$ Universe$) \to \mathcal{U} \sqcup \mathcal{W} \sqcup \mathcal{X}$ $^+$ ˙
REL $A$ $B$ $\mathcal{X} = A \to B \to \mathcal{X}$ ˙

Rel : $\mathcal{U}$ ˙ $\to (\mathcal{X} :$ Universe$) \to \mathcal{U} \sqcup \mathcal{X}$ $^+$ ˙
Rel $A$ $\mathcal{X} =$ REL $A$ $A$ $\mathcal{X}$

### The kernel of a function

The *kernel* of a function $f : A \to B$ is defined informally by $\{(x, y) \in A \times A : f\, x = f\, y\}$. This can be represented in type theory in a number of ways, each of which may be useful in a particular context. For example, we could define the kernel to be an inhabitant of a (binary) relation type, a (unary) predicate type, a (curried) Sigma type, or an (uncurried) Sigma type. Since the first two alternatives are the ones we use thoughout the UALib, we present them here.

ker : $(A \to B) \to$ Rel $A$ $\mathcal{W}$
ker $g\, x\, y = g\, x \equiv g\, y$

kernel : $(A \to B) \to$ Pred $(A \times A)$ $\mathcal{W}$
kernel $g\, (x, y) = g\, x \equiv g\, y$

Similarly, the *identity relation* (which is equivalent to the kernel of an injective function) can be represented by a number of different types. Here we only show the representation that we use later to construct the zero congruence. The notation we use here is close to that of conventional algebra notation, where $0_A$ is used to denote the identity relation $\{(x,y) \in A \times A : x = y\}$.[25]

𝟎 : Rel $A$ $\mathcal{U}$
𝟎 $x\, y = x \equiv y$

### The implication relation[27]

The following types represent *implication* for binary relations.

---

[27] The definitions here are from the Agda Standard Library, translated into TypeTopo/UALib notation.

```
_on_ : (B → B → C) → (A → B) → (A → A → C)
R on g = λ x y → R (g x) (g y)

_⇒_ : REL A B 𝒳 → REL A B 𝒴 → 𝒰 ⊔ 𝒲 ⊔ 𝒳 ⊔ 𝒴 ˙
P ⇒ Q = ∀ {i j} → P i j → Q i j
```

These combine to give a nice, general implication operation.

```
_=[_]⇒_ : Rel A 𝒳 → (A → B) → Rel B 𝒴 → 𝒰 ⊔ 𝒳 ⊔ 𝒴 ˙
P =[ g ]⇒ Q = P ⇒ (Q on g)
```

## Operation type and compatibility

**Notation**. For consistency and readability, throughout the UALib we reserve two universe variables for special purposes. The first of these is 𝒪 which shall be reserved for types that represent *operation symbols* (see Algebras.Signatures). The second is 𝒱 which we reserve for types representing *arities* of relations or operations.

Below we will define types that are useful for asserting and proving facts about *compatibility* of operations and relations, but first we need a general type with which to represent operations. Here is the definition, which we justify below.

```
Op : 𝒱 ˙ → 𝒰 ˙ → 𝒰 ⊔ 𝒱 ˙
Op I A = (I → A) → A
```

The definition of Op codifies the arity of an operation as an arbitrary type $I : \mathcal{V}$ ˙, which gives us a very general way to represent an operation as a function type with domain $I \to A$ (the type of "*I*-tuples") and codomain $A$. For example, the *I*-ary projection operations on $A$ are represented as inhabitants of the type Op $I$ $A$ as follows.

```
π : {I : 𝒱 ˙ } {A : 𝒰 ˙ } → I → Op I A
π i x = x i
```

Let us review the informal definition of compatibility. Suppose $A$ and $I$ are types and fix $f$ : Op $I$ $A$ and $R$ : Rel $A$ 𝒲 (an *I*-ary operation and a binary relation on $A$, respectively). We say that $f$ and $R$ are *compatible* and we write[28] $f$ |: $R$ just in case $\forall\ u\ v : I \to A$,

Π $i : I$, $R\ (u\ i)\ (v\ i)$ → $R\ (f\ u)\ (f\ v)$.

To implement this in Agda, we first define a function eval-rel which "lifts" a binary relation to the corresponding *I*-ary relation, and we use this to define the function |: representing *compatibility of an I-ary operation with a binary relation*.

```
eval-rel : {A : 𝒰 ˙}{I : 𝒱 ˙} → Rel A 𝒲 → Rel (I → A)(𝒱 ⊔ 𝒲)
eval-rel R u v = Π i : _ , R (u i) (v i)

_|:_ : {A : 𝒰 ˙}{I : 𝒱 ˙} → Op I A → Rel A 𝒲 → 𝒰 ⊔ 𝒱 ⊔ 𝒲 ˙
f |: R = (eval-rel R) =[ f ]⇒ R
```

---

[28] The symbol |: denoting compatibility comes from Cliff Bergman's universal algebra textbook [1].

### 3.2 Continuous relations*[29]

This section presents the Relations.Continuous module of the AgdaUALib.

**Motivation**

In set theory, an *n*-ary relation on a set $A$ is a subset of the *n*-fold product $A \times \cdots \times A$. We could try to model these as predicates over a product type representing $A \times \cdots \times A$, or as relations of type $A \to A \to \cdots \to A \to \mathcal{W}$ ˙ (for some universe $\mathcal{W}$). To implement such types requires knowing the arity in advance, and then form an *n*-fold product or *n*-fold arrow. It turns out to be both easier and more general if we define an arity type $I : \mathcal{V}$ ˙ , and define the type representing *I*-ary relations on $A$ as the function type $(I \to A) \to \mathcal{W}$ ˙. Then, if we happen to be interested in *n*-ary relations for some natural number *n*, we could take $I$ to be the *n*-element type Fin *n*, [23].

Below we will define ContRel to be the type $(I \to A) \to \mathcal{W}$ ˙ and we will call this the type of *continuous relations*. This generalizes the discrete relations we defined in Relations.Discrete (unary, binary, etc.) since continuous relations can be of arbitrary arity. They are not completely general, however, since they are defined over a single type. Said another way, these are "single-sorted" relations. We will remove this limitation when we define the type of *dependent continuous relations*. Just as Rel $A$ $\mathcal{W}$ was the single-sorted special case of the multisorted REL $A$ $B$ $\mathcal{W}$ type, so too will ContRel $I$ $A$ $\mathcal{W}$ be the single-sorted version of dependent continuous relations. The latter will represent relations that not only have arbitrary arities, but also are defined over arbitrary families of types.

To be more concrete, given an arbitrary family $A : I \to \mathcal{U}$ ˙ of types, we may have a relation from $A$ *i* to $A$ *j* to $A$ *k* to ..., where the collection represented by the indexing type $I$ might not even be enumerable.[30] We will refer to such relations as *dependent continuous relations* (or *dependent relations*) because the definition of a type that represents them requires dependent types. The DepRel type that we define below manifests this completely general notion of relation.

**Continuous and dependent relation types**

Here we define the types ContRel and DepRel. The first of these represents predicates of arbitrary arity over a single type $A$; we call these *continuous relations*.[31] To define the type DepRel of *dependent relations*, we exploit the full power of dependent types to provide a completely general relation type.

ContRel : $\mathcal{V}$ ˙ → $\mathcal{U}$ ˙ → ($\mathcal{W}$ : Universe) → $\mathcal{V}$ ⊔ $\mathcal{U}$ ⊔ $\mathcal{W}$ ⁺ ˙
ContRel $I$ $A$ $\mathcal{W}$ = $(I \to A) \to \mathcal{W}$ ˙

DepRel : $(I : \mathcal{V}$ ˙) → $(I \to \mathcal{U}$ ˙) → ($\mathcal{W}$ : Universe) → $\mathcal{V}$ ⊔ $\mathcal{U}$ ⊔ $\mathcal{W}$ ⁺ ˙
DepRel $I$ $\mathcal{A}$ $\mathcal{W}$ = Π $\mathcal{A}$ → $\mathcal{W}$ ˙

---

[29] Sections marked with an asterisk may be safely skimmed or skipped on a first reading; they present some potentially interesting new types that are not yet employed in other parts of the UALib.

[30] Because the collection represented by the indexing type $I$ might not even be enumerable, technically speaking, instead of "$A$ *i* to $A$ *j* to $A$ *k* to ...," we should have written something like "TO $(i : I)$, $A$ *i*."

[31] For consistency and readability, throughout the UALib we reserve two universe variables for special purposes: $\mathcal{O}$ is reserved for types representing *operation symbols*; $\mathcal{V}$ is reserved for types representing *arities* of relations or operations (see Algebras.Signatures).

Here, the tuples of a relation of type DepRel $I$ 𝒜 𝒲 will inhabit the dependent function type 𝒜 : $I \to$ 𝒰 ˙, where the codomain may depend on the input coordinate $i : I$ of the domain. Heuristically, we can think of an inhabitant of DepRel $I$ 𝒜 𝒲 as a relation from 𝒜 $i$ to 𝒜 $j$ to 𝒜 $k$ to …. (This is only a rough heuristic since $I$ could denote an uncountable collection.[30]

### Compatibility with general relations

It will be helpful to have functions that make it easy to assert that a given operation is compatibile with a given relation. The following functions serve this purpose.

    eval-cont-rel : ContRel $I$ $A$ 𝒲 $\to (I \to J \to A) \to$ 𝒱 ⊔ 𝒲 ˙
    eval-cont-rel $R$ 𝒶 $= \Pi\ j : J\ ,\ R\ \lambda\ i \to$ 𝒶 $i\ j$

    cont-compatible-op : Op $J$ $A \to$ ContRel $I$ $A$ 𝒲 $\to$ 𝒱 ⊔ 𝒰 ⊔ 𝒲 ˙
    cont-compatible-op $f\ R = \Pi$ 𝒶 $: (I \to J \to A)\ ,$ (eval-cont-rel $R$ 𝒶 $\to R\ \lambda\ i \to (f\ ($𝒶 $i)))$

The first of these is an *evaluation* function which "lifts" an $I$-ary relation to an $(I \to J)$-ary relation. The lifted relation will relate an $I$-tuple of $J$-tuples when the "$I$-slices" (or "rows") of the $J$-tuples belong to the original relation. The second definition denotes compatibility of an operation with a continuous relation.

Readers who find the syntax of the last two definitions nauseating might be helped by an explication of the semantics of these definitions. First, internalize the fact that 𝒶 $: I \to J \to A$ denotes an $I$-tuple of $J$-tuples of inhabitants of $A$. Next, recall that a continuous relation $R$ denotes a certain collection of $I$-tuples (if $x : I \to A$, then $R\ x$ asserts that $x$ "belongs to" or "satisfies" $R$). For such $R$, the type eval-cont-rel $R$ represents a certain collection of $I$-tuples of $J$-tuples, namely, the tuples 𝒶 $: I \to J \to A$ for which eval-cont-rel $R$ 𝒶 holds.

For simplicity, pretend for a moment that $J$ is a finite set, say, $\{1, 2, ..., J\}$, so that we can write down a couple of the $J$-tuples as columns. For example, here are the $i$-th and $k$-th columns (for some $i\ k : I$).

    𝒶 $i$ 1    𝒶 $k$ 1
    𝒶 $i$ 2    𝒶 $k$ 2    ← (if there are $I$ columns, then each row forms an $I$-tuple)
      ⋮     ⋮
    𝒶 $i$ $J$    𝒶 $k$ $J$

Now eval-cont-rel $R$ 𝒶 is defined by $\forall\ j \to R\ (\lambda\ i \to ($𝒶 $i)\ j)$ which asserts that each row of the $I$ columns shown above belongs to the original relation $R$. Finally, cont-compatible-op takes a $J$-ary operation on $A$, say, $f :$ Op $J$ $A$, and an $I$-tuple 𝒶 $i : J \to A$ of $J$-tuples, and determines whether the $I$-tuple $\lambda\ i \to f\ ($𝒶 $i)$ belongs to $R$.

We conclude this section by defining the (only slightly more complicated) lift of dependent relations, and the type that represents compatibility of a tuple of operations with a dependent relation. Here we assume $I\ J :$ 𝒱 ˙ and 𝒜 $: I \to$ 𝒰 ˙.

    eval-dep-rel : DepRel $I$ 𝒜 𝒲 $\to (\forall\ i \to (J \to$ 𝒜 $i)) \to$ 𝒱 ⊔ 𝒲 ˙
    eval-dep-rel $R$ 𝒶 $= \forall\ j \to R\ (\lambda\ i \to ($𝒶 $i)\ j)$

    dep-compatible-op : $(\forall\ i \to$ Op $J$ $($𝒜 $i)) \to$ DepRel $I$ 𝒜 𝒲 $\to$ 𝒱 ⊔ 𝒰 ⊔ 𝒲 ˙
    dep-compatible-op $f\ R = \forall$ 𝒶 $\to$ (eval-dep-rel $R$) 𝒶 $\to R\ \lambda\ i \to (f\ i)($𝒶 $i)$

where we let Agda infer that the type of 𝒶 is $\Pi\ i : I\ ,\ (J \to$ 𝒜 $i)$.

## 3.3 Equivalence relations and quotients

This section presents the Relations.Quotients module of the AgdaUALib, slightly abridged.[32]

**Equivalence relations**

In the Relations.Discrete module we defined types for representing and reasoning about binary relations on $A$. In this module we will define types for binary relations that have special properties. The most important special properties of relations are the ones we now define.

>  Refl : $\{A : \mathcal{U} \ \cdot\} \to$ Rel $A \ \mathcal{W} \to \mathcal{U} \sqcup \mathcal{W} \ \cdot$
>  Refl $\_\approx\_ = \forall\{x\} \to x \approx x$
>
>  Symm : $\{A : \mathcal{U} \ \cdot\} \to$ Rel $A \ \mathcal{W} \to \mathcal{U} \sqcup \mathcal{W} \ \cdot$
>  Symm $\_\approx\_ = \forall\{x\}\{y\} \to x \approx y \to y \approx x$
>
>  Antisymm : $\{A : \mathcal{U} \ \cdot\} \to$ Rel $A \ \mathcal{W} \to \mathcal{U} \sqcup \mathcal{W} \ \cdot$
>  Antisymm $\_\approx\_ = \forall\{x\}\{y\} \to x \approx y \to y \approx x \to x \equiv y$
>
>  Trans : $\{A : \mathcal{U} \ \cdot\} \to$ Rel $A \ \mathcal{W} \to \mathcal{U} \sqcup \mathcal{W} \ \cdot$
>  Trans $\_\approx\_ = \forall\{x\}\{y\}\{z\} \to x \approx y \to y \approx z \to x \approx z$

Here is a type from TypeTopo that expresses *proof-irrelevance* for binary relations.

>  is-subsingleton-valued : $\{A : \mathcal{U} \ \cdot\} \to$ Rel $A \ \mathcal{W} \to \mathcal{U} \sqcup \mathcal{W} \ \cdot$
>  is-subsingleton-valued $\_\approx\_ = \forall \ x \ y \to$ is-subsingleton $(x \approx y)$

Thus, if $R$ : Rel $A \ \mathcal{W}$, then is-subsingleton-valued $R$ asserts that for each pair $x \ y : A$ there is *at most one proof* of $R \ x \ y$. In the section on *Truncation* below (§ 3.4) we introduce a number of similar but more general types to represent *uniqueness-of-proofs* principles for relations of arbitrary arity, over arbitrary types.

A binary relation is called a *preorder* if it is reflexive and transitive, and an *equivalence relation* is a symmetric preorder. We represent the property of being an equivalence relation by the following record type.

>  record IsEquivalence $\{A : \mathcal{U} \ \cdot\}(R :$ Rel $A \ \mathcal{W}) : \mathcal{U} \sqcup \mathcal{W} \ \cdot$ where
>    field rfl : Refl $R$; sym : Symm $R$; trans : Trans $R$

The type of equivalence relations is then defined as follows.

>  Equivalence : $\mathcal{U} \ \cdot \to \mathcal{U} \sqcup \mathcal{W}^{+} \ \cdot$
>  Equivalence $A = \Sigma \ R :$ Rel $A \ \mathcal{W}$ , IsEquivalence $R$

Thus, if we have $(R , p)$ : Equivalence $A$, then $R$ denotes a binary relation over $A$ and $p$ is of record type IsEquivalence $R$ whose fields contain the three proofs required to show that $R$ is an equivalence relation.

A prominent example of an equivalence relation is the kernel of any function.

>  ker-IsEquivalence : $\{A : \mathcal{U} \ \cdot\}\{B : \mathcal{W} \ \cdot\}(f : A \to B) \to$ IsEquivalence (ker $f$)
>  ker-IsEquivalence $f =$ record { rfl = refl; sym = $\lambda \ z \to \equiv$-sym $z$ ; trans = $\lambda \ p \ q \to \equiv$-trans $p \ q$ }

---

[32] For unabridged docs (source code) see https://ualib.gitlab.io/Relations.Quotients.html (https://gitlab.com/ualib/ualib.gitlab.io/-/blob/master/UALib/Relations/Quotients.lagda).

**Quotients**

If $R$ is an equivalence relation on $A$, then for each $u : A$ there is an *equivalence class* (or *equivalence block*, or *R-block*) containing $u$, which we denote by $[\ u\ ] := \{v : A \mid R\ u\ v\}$. We call this the *R-block containing $a$*.

    [\_] : $\{A : \mathcal{U}\ \cdot\} \to A \to \{R : $ Rel $A\ \mathcal{W}\} \to$ Pred $A\ \mathcal{W}$
    [ $u$ ]$\{R\} = R\ u$

Thus, $v \in [\ u\ ]$ if and only if $R\ u\ v$, as desired. We often refer to $[\ u\ ]$ as the *R-block containing $u$*. A predicate $C$ over $A$ is an $R$-block if and only if $C \equiv [\ u\ ]$ for some $u : A$. We represent this characterization of an $R$-block as follows.

    IsBlock : $\{A : \mathcal{U}\ \cdot\}(C :$ Pred $A\ \mathcal{W})\{R :$ Rel $A\ \mathcal{W}\} \to \mathcal{U} \sqcup \mathcal{W}\ ^+\ \cdot$
    IsBlock $\{A\}\ C\ \{R\} = \Sigma\ a : A\ ,\ C \equiv [\ a\ ]\ R$

Thus, a proof of the assertion IsBlock $C$ is a dependent pair $(u\ ,\ p)$, with $u : A$ and $p$ is a proof of $C \equiv [\ u\ ]\{R\}$.

If $R$ is an equivalence relation on $A$, then the *quotient of $A$ modulo $R$* is denoted by $A\ /\ R$ and is defined to be the collection $\{[\ u\ ] \mid y : A\}$ of all $R$-blocks.

    \_/\_ : $(A : \mathcal{U}\ \cdot\ ) \to$ Rel $A\ \mathcal{W} \to \mathcal{U} \sqcup (\mathcal{W}\ ^+)\ \cdot$
    $A\ /\ R = \Sigma\ C :$ Pred $A\ \mathcal{W}\ ,$ IsBlock $C\ \{R\}$

We use the following type to represent an $R$-block with a designated representative.

    ⟪\_⟫ : $\{A : \mathcal{U}\ \cdot\} \to A \to \{R :$ Rel $A\ \mathcal{W}\} \to A\ /\ R$
    ⟪ $a$ ⟫ $\{R\} = [\ a\ ]\ R\ ,\ (a\ ,$ refl$)$

This serves as a kind of *introduction rule*. Dually, the next type provides an *elimination rule*.[33]

    ⌞\_⌟ : $\{A : \mathcal{U}\ \cdot\}\{R :$ Rel $A\ \mathcal{W}\} \to A\ /\ R \to A$
    ⌞ $C\ ,\ (a\ ,\ p)$ ⌟ $= a$

It will be convenient to have the following subset inclusion lemmas on hand when working with quotient types. (Assume a context including $\mathcal{U}\ \mathcal{W} :$ Universe, $A : \mathcal{U}\ \cdot$, $x\ y :\ A$, and $R :$ Rel $A\ \mathcal{W}$.)

    /-subset : IsEquivalence $R \to R\ x\ y \to [\ x\ ]\ R \subseteq [\ y\ ]\ R$
    /-subset $Req\ Rxy\ \{z\}\ Rxz = ($trans $Req)$ $(($sym $Req)\ Rxy)\ Rxz$

    /-supset : IsEquivalence $R \to R\ x\ y \to [\ y\ ]\ R \subseteq [\ x\ ]\ R$
    /-supset $Req\ Rxy\ \{z\}\ Ryz = ($trans $Req)\ Rxy\ Ryz$

An example application of these is the block-ext type in the Relations.Extensionality module.

### 3.4 Truncation

This section presents the Relations.Truncation module of the AgdaUALib, slightly abridged.[34] We start with a brief discussion of standard notions of *truncation*, *h-sets* (which we just call *sets*), and the *uniqueness of identity types* principle. We then prove that a monic function into a *set* is an embedding. We conclude the section with a *uniqueness of identity proofs* principle for blocks of equivalence relations.[35]

---

[33] **Unicode Hint**. Type ⌞ and ⌟ as \cul and \cur in agda2-mode.
[34] For unabridged docs (source code) see https://ualib.gitlab.io/Relations.Truncation.html (https://gitlab.com/ualib/ualib.gitlab.io/-/blob/master/UALib/Relations/Truncation.lagda).

### Uniqueness of identity proofs

This brief introduction is intended for novices; those already familiar with the concept of *truncation* and *uniqueness of identity proofs* (UIP) may want to skip to the next subsection.

In general, we may have multiple inhabitants of a given type, hence (via Curry-Howard) multiple proofs of a given proposition. For instance, suppose we have a type $X$ and an identity relation $\_\equiv_0\_$ on $X$ so that, given two inhabitants of $X$, say, $a\ b : X$, we can form the type $a \equiv_0 b$. Suppose $p$ and $q$ inhabit the type $a \equiv_0 b$; that is, $p$ and $q$ are proofs of $a \equiv_0 b$, in which case we write $p\ q : a \equiv_0 b$. We might then wonder whether the two proofs $p$ and $q$ are equivalent.

We are asking about an identity type on the identity type $\equiv_0$, and whether there is some inhabitant, say, $r$ of this type; i.e., whether there is a proof $r : p \equiv_1 q$ that the proofs of $a \equiv_0 b$ are the same. If such a proof exists for all $p\ q : a \equiv_0 b$, then the proof of $a \equiv_0 b$ is unique; as a property of the types $X$ and $\equiv_0$, this is sometimes called *uniqueness of identity proofs* (UIP).

Now, perhaps we have two proofs, say, $r\ s : p \equiv_1 q$ that the proofs $p$ and $q$ are equivalent. Then of course we wonder whether $r \equiv_2 s$ has a proof! But at some level we may decide that the potential to distinguish two proofs of an identity in a meaningful way (so-called *proof-relevance*) is not useful or desirable. At that point, say, at level $k$, we would be naturally inclined to assume that there is at most one proof of any identity of the form $p \equiv_k q$. This is called truncation (at level $k$).

### Sets

In homotopy type theory, a type $X$ with an identity relation $\equiv_0$ is called a *set* (or *0-groupoid*) if for every pair $x\ y : X$ there is at most one proof of $x \equiv_0 y$. In other words, the type $X$, along with it's equality type $\equiv_x$, form a *set* if for all $x\ y : X$ there is at most one proof of $x \equiv_0 y$. This notion is formalized in TypeTopo, using the is-subsingleton type (§2.4), as follows.

```
is-set : 𝒰 ˙ → 𝒰 ˙
is-set A = (x y : A) → is-subsingleton (x ≡ y)
```

Thus, the pair $(X, \equiv_0)$ forms a set iff it satisfies $\forall\ x\ y : X \rightarrow$ is-subsingleton $(x \equiv_0 y)$.

We will also need the type to-$\Sigma$-$\equiv$, part of Escardó's characterization of *equality in Sigma types* in [10], defined as follows. (Assume a context including $A : \mathcal{U}\ \dot{}$ and $B : A \rightarrow \mathcal{W}\ \dot{}$.)

```
to-Σ-≡ : {σ τ : Σ B} → Σ p : | σ | ≡ | τ | , (transport B p ∥ σ ∥) ≡ ∥ τ ∥ → σ ≡ τ
to-Σ-≡ (refl {x = x} , refl {x = a}) = refl {x = (x , a)}
```

### Injective functions are set embeddings

Before moving on to define propositions, we discharge an obligation mentioned but left unfulfilled in the embeddings section of the Overture.Inverses module. Recall, we described and imported the is-embedding type, and we remarked that an embedding is not simply a monic function. However, if we assume that the codomain is truncated so as to have unique identity proofs, then we can prove that every monic function into that codomain will be an embedding.

---

[35] Readers who want to learn more about "proof-relevant mathematics" and other concepts mentioned in this section may wish to consult other sources, such as [10, §34], [3], or [18, §7.1].

On the other hand, embeddings are always monic, so we will end up with an equivalence.

```
monic-is-embedding|Set : (f : A → B) → is-set B → Monic f → is-embedding f
monic-is-embedding|Set f Bset fmon b (u , fu≡b) (v , fv≡b) = γ
  where
  fuv : f u ≡ f v
  fuv = ≡-trans fu≡b (fv≡b ⁻¹)

  uv : u ≡ v
  uv = fmon u v fuv

  arg1 : Σ p : (u ≡ v) , (transport (λ a → f a ≡ b) p fu≡b) ≡ fv≡b
  arg1 = uv , Bset (f v) b (transport (λ a → f a ≡ b) uv fu≡b) fv≡b

  γ : u , fu≡b ≡ v , fv≡b
  γ = to-Σ-≡ arg1
```

In stating the previous result, we introduce a new convention to which we will try to adhere. If the antecedent of a theorem includes the assumption that one of the types involved is a set, then we add to the name of the theorem the suffix |sets, which calls to mind the standard notation for the restriction of a function to a subset of its domain.

### Equivalence class truncation

Recall, the definition IsBlock $C = \Sigma\ u : A\ ,\ C \equiv [\ u\ ]$, from the Relations.Quotients module. In the next module (Relations.Extensionality) we will define a *quotient extensionality* principle that will require a form of unique identity proofs—specifically, we will assume that for each predicate $C$ : Pred $A\ \mathcal{W}$ there is at most one proof of IsBlock $C$. We call this proof-irrelevance principle "uniqueness of block identity proofs" and define it as follows.

```
blk-uip : {𝓦 𝓤 : Universe}(A : 𝓤 ˙)(R : Rel A 𝓦 ) → 𝓤 ⊔ 𝓦 ⁺ ˙
blk-uip {𝓦} A R = ∀ (C : Pred A 𝓦) → is-subsingleton (IsBlock C {R})
```

It might seem unreasonable to postulate that there is at most one inhabitant of IsBlock $C$, since equivalence classes typically have multiple members, any one of which could serve as a class representative. However, postulating blk-uip $A\ R$ is tantamount to collapsing each $R$-block to a single point, and this is indeed the correct semantic interpretation of the elements of the quotient $A\ /\ R$.

### General propositions*[29]

This section defines more general truncated predicates which we call *continuous propositions* and *dependent propositions*. Recall, above (in the Relations.Continuous module) we defined types called ContRel and DepRel to represent relations of arbitrary arity over arbitrary collections of sorts. Naturally, we define the corresponding *truncated continuous relation type* and *truncated dependent relation type*, the inhabitants of which we will call *continuous propositions* and *dependent propositions*, respectively. (Assume the context includes $\mathcal{U}$ : Universe and $I : \mathcal{V}$ ˙.)

```
IsContProp : {A : 𝓤 ˙}{𝓦 : Universe} → ContRel I A 𝓦 → 𝓥 ⊔ 𝓤 ⊔ 𝓦 ˙
IsContProp {A = A} P = Π a : (I → A) , is-subsingleton (P a)

ContProp : 𝓤 ˙ → (𝓦 : Universe) → 𝓤 ⊔ 𝓥 ⊔ 𝓦 ⁺ ˙
ContProp A 𝓦 = Σ P : (ContRel I A 𝓦) , IsContProp P
```

```
cont-prop-ext : 𝒰 ˙ → (𝒲 : Universe) → 𝒰 ⊔ 𝒱 ⊔ 𝒲 ⁺ ˙
cont-prop-ext A 𝒲 = {P Q : ContProp A 𝒲 } → | P | ⊆ | Q | → | Q | ⊆ | P | → P ≡ Q

IsDepProp : {I : 𝒱 ˙}{𝒜 : I → 𝒰 ˙}{𝒲 : Universe} → DepRel I 𝒜 𝒲 → 𝒱 ⊔ 𝒰 ⊔ 𝒲 ˙
IsDepProp {I = I} {𝒜} P = Π a : Π 𝒜 , is-subsingleton (P a)

DepProp : (I → 𝒰 ˙) → (𝒲 : Universe) → 𝒰 ⊔ 𝒱 ⊔ 𝒲 ⁺ ˙
DepProp 𝒜 𝒲 = Σ P : (DepRel I 𝒜 𝒲) , IsDepProp P

dep-prop-ext : (I → 𝒰 ˙) → (𝒲 : Universe) → 𝒰 ⊔ 𝒱 ⊔ 𝒲 ⁺ ˙
dep-prop-ext 𝒜 𝒲 = {P Q : DepProp 𝒜 𝒲} → | P | ⊆ | Q | → | Q | ⊆ | P | → P ≡ Q
```

To see the point of the types just defined, suppose dep-prop-ext $A$ 𝒲 holds. Then we can prove that logically equivalent dependent propositions (of type DepProp $A$ 𝒲) are equivalent. In other words, under the stated hypotheses, we obtain the following extensionality lemma for dependent propositions (assuming the context includes 𝒜 : $I$ → 𝒰 ˙ and 𝒲 : Universe).

```
dep-prop-ext' : dep-prop-ext 𝒜 𝒲 → {P Q : DepProp 𝒜 𝒲} → | P | ≐ | Q | → P ≡ Q
dep-prop-ext' pe hyp = pe | hyp | ∥ hyp ∥
```

## 3.5 Relation extensionality

This section presents the Relations.Extensionality module of the AgdaUALib, slightly abridged.[36]

The principle of *proposition extensionality* asserts that logically equivalent propositions are equivalent. That is, if $P$ and $Q$ are propositions such that $P ⊆ Q$ and $Q ⊆ P$, then $P ≡ Q$. For our purposes, it will suffice to formalize this principle for general predicates, rather than propositions (i.e., truncated predicates). The following definition codifies the extensionality principle we require.

```
pred-ext : (𝒰 𝒲 : Universe) → (𝒰 ⊔ 𝒲) ⁺ ˙
pred-ext 𝒰 𝒲 = ∀ {A : 𝒰 ˙}{P Q : Pred A 𝒲 } → P ⊆ Q → Q ⊆ P → P ≡ Q
```

Note that pred-ext does not itself postulate the axiom of predicate extensionality; it merely defines what that axiom asserts. If we want to postulate the axiom, we must assume we have an inhabitant (or "witness") of the pred-ext type (see block-ext below, for example).

### Quotient extensionality

We need an identity type for congruence classes (blocks) over sets so that two different presentations of the same block (e.g., using different representatives) may be identified. This requires two postulates: (1) *predicate extensionality* manifested by the blk-uip type defined earlier (§3.4). We now use pred-ext and blk-uip to define a type called block-ext|uip which we require for the proof of the *First Homomorphism Theorem* presented in Homomorphisms.Noether. (Assume a context including 𝒰 𝒲 : Universe, $A$ : 𝒰 ˙, and $R$ : Rel $A$ 𝒲.)

```
block-ext : pred-ext 𝒰 𝒲 → IsEquivalence R → {u v : A} → R u v → [ u ]{R} ≡ [ v ]{R}
block-ext pe Req {u}{v} Ruv = pe (/-subset Req Ruv) (/-supset Req Ruv)
```

---

[36] For unabridged docs (source code) see https://ualib.gitlab.io/Relations.Extensionality.html (https://gitlab.com/ualib/ualib.gitlab.io/-/blob/master/UALib/Relations/Extensionality.lagda).

```
to-subtype|uip : blk-uip A R → {C D : Pred A 𝒲}{c : IsBlock C {R}}{d : IsBlock D {R}}
  → C ≡ D → (C , c) ≡ (D , d)
to-subtype|uip buip {C}{D}{c}{d} CD = to-Σ-≡(CD , buip D(transport(λ B → IsBlock B)CD c)d)

block-ext|uip : pred-ext 𝒰 𝒲 → blk-uip A R → IsEquivalence R
  → {u v : A} → R u v → ⟪ u ⟫ ≡ ⟪ v ⟫
block-ext|uip pe buip Req Ruv = to-subtype|uip buip (block-ext pe Req Ruv)
```

## 4   Algebra Types

Finally we are ready to present the Algebras module of the AgdaUALib. Here we use type theory and Agda to codify the most basic objects of universal algebra, such as *operations* and *signatures* (§4.1), *algebras* (§4.2), *product algebras* (§4.3), *congruence relations* and *quotient algebras* (§4.4).

A popular way to represent algebraic structures in type theory is with record types. The Sigma type provides an equivalent alternative that we happen to prefer and we use it throughout the library, both for consistency and because of its direct connection to the existential quantifier of logic. Recall that inhabitants of the type $\Sigma\ x : X\ ,\ P$ are pairs $(x, p)$ such that $x : X$ and $p : P\ x$. In this sense, when such a Sigma type is inhabited we conclude, "there exists $x$ in $X$ such that $P\ x$ holds;" in symbols, $\exists\ x \in X\ ,\ P\ x$. Moreover, the pair $(x, p)$ is not merely a proof of the logical sentence $\exists\ x \in X\ ,\ P\ x$; it is also a *witness* of the truth of this sentence. We sometimes say that a proof of an existentially quantified sentence has "computational content" if it provides such a witness, or a function that can extract a witness from the proof.

### 4.1   Signatures

This section presents the Algebras.Signatures module of the AgdaUALib, slightly abridged.[37]

We define the signature of an algebraic structure in Agda like this.

```
Signature : (𝓞 𝒱 : Universe) → (𝓞 ⊔ 𝒱) ⁺ ˙
Signature 𝓞 𝒱 = Σ F : 𝓞 ˙ , (F → 𝒱 ˙)
```

As mentioned in §3.1, the symbol 𝓞 always denotes the universe of *operation symbol* types, while 𝒱 is always the universe of *arity* types.

In §2.1 we defined special syntax for the first and second projections—namely, |_| and ||_||, respectively. Consequently, if $S$ : Signature 𝓞 𝒱 is a signature, then $\mid S \mid$ denotes the set of operation symbols, and $\parallel S \parallel$ denotes the arity function. If $f : \mid S \mid$ is an operation symbol in the signature $S$, then $\parallel S \parallel\ f$ is the arity of $f$.

**Example of a signature**

Here is how we could define the signature for *monoids* as an inhabitant of the type Signature 𝓞 𝒱.

```
data monoid-op : 𝓞 ˙ where
  e : monoid-op
  · : monoid-op

monoid-sig : Signature 𝓞 𝒰₀
monoid-sig = monoid-op , λ { e → 𝟘; · → 𝟚 }
```

---

[37] For unabridged docs (source code) see https://ualib.gitlab.io/Algebras.Signatures.html (https://gitlab.com/ualib/ualib.gitlab.io/-/blob/master/UALib/Algebras/Signatures.lagda).

Thus, the signature for a monoid consists of two operation symbols, e and ·, and a function $\lambda \{ \text{e} \to \mathbb{0} ; \cdot \to \mathbb{2} \}$ which maps e to the empty type $\mathbb{0}$ (since e is the nullary identity) and maps · to the two element type $\mathbb{2}$ (since · is binary).[38]

## 4.2 Algebras

This section presents the Algebras.Algebras module of the AgdaUALib, slightly abridged.[39]

Recall, the signature type Signature $\mathcal{O}$ $\mathcal{V}$ (§4.1) was defined as the Sigma type $\Sigma F : \mathcal{O} \cdot , (F \to \mathcal{V} \cdot )$. The operation symbol type Op $I$ $A$ (§3.1) was defined as the function type $(I \to A) \to A$. For a fixed signature $S$ : Signature $\mathcal{O}$ $\mathcal{V}$ and universe $\mathcal{U}$, we define the type of *algebras in the signature $S$* (or *$S$-algebras*) with *domain* (or *carrier*) of type $\mathcal{U} \cdot$ as follows.[40]

```
Algebra : (𝒰 : Universe)(S : Signature 𝒪 𝒱) → 𝒪 ⊔ 𝒱 ⊔ 𝒰 ⁺ ·
Algebra 𝒰 S = Σ A : 𝒰 · ,              – the domain
              Π f : | S | , Op (∥ S ∥ f) A   – the basic operations
```

To be precise we could call an inhabitant of this type an "∞-algebra" because its domain can be an arbitrary type, say, $A : \mathcal{U} \cdot$ and need not be truncated at some level. In particular, $A$ need not be a *set* (§3.4). We could then proceed to define the type of "0-algebras" as algebras whose domains would be sets, which may be closer to what most of us have in mind when doing informal universal algebra. However, we have found that the domains of our algebras need to be sets in just a few places in the UALib, and it seems preferable to work with general (∞-)algebras throughout and then assume *uniqueness of identity proofs* (UIP) explicitly and only where needed. This makes any dependence on UIP more transparent (which is also the reason *–without-K* appears at the top of every module in the UALib).

**Operation interpretation syntax**

We now define a convenient shorthand for the interpretation of an operation symbol. This looks more similar to the standard notation one finds in the literature as compared to the double bar notation we started with, so we will use this new notation almost exclusively in the remaining modules of the UALib.

```
_^_ : (f : | S |)(A : Algebra 𝒰 S) → (∥ S ∥ f → | A |) → | A |

f ^ A = λ a → (∥ A ∥ f) a
```

Thus, if $f : | S |$ is an operation symbol in the signature $S$ and if $a : \| S \| f \to | \mathbf{A} |$ is a tuple of the same arity, then $(f \; \hat{} \; \mathbf{A}) \; a$ denotes the operation $f$ interpreted in $\mathbf{A}$ and evaluated at $a$.

**Lifts of algebras**

Recall, in §2.5 we described a common difficulty one encounters when working with a noncumulative universe hierarchy. We made a promise to provide some domain-specific level lifting and lowering methods. Here we fulfill this promise by supplying a couple of bespoke tools designed

---

[38] The types $\mathbb{0}$ and $\mathbb{2}$ are defined in the MGS-MLTT module of TypeTopo.

[39] For unabridged docs (source code) see https://ualib.gitlab.io/Algebras.Algebras.html (https://gitlab.com/ualib/ualib.gitlab.io/-/blob/master/UALib/Algebras/Algebras.lagda).

[40] Universal algebraists often call the domain of an algebra its *universe*. We avoid this terminology since "universe" is used in type theory for levels of the type hierarchy.

specifically for our operation and algebra types.

    Lift-op : $((I \to A) \to A) \to (\mathcal{W} :$ Universe$) \to ((I \to$ Lift$\{\mathcal{W}\}\ A) \to$ Lift $\{\mathcal{W}\}\ A)$
    Lift-op $f\ \mathcal{W} = \lambda\ x \to$ lift $(f\ (\lambda\ i \to$ lower $(x\ i)))$

    Lift-alg : Algebra $\mathcal{U}\ S \to (\mathcal{W} :$ Universe$) \to$ Algebra $(\mathcal{U} \sqcup \mathcal{W})\ S$
    Lift-alg **A** $\mathcal{W} =$ Lift $\mid$ **A** $\mid$ , $(\lambda\ (f :\mid S \mid) \to$ Lift-op $(f\ \hat{}\ $**A**$)\ \mathcal{W})$

What makes the Lift-alg type so useful for resolving type level errors is the nice properties it possesses. Indeed, in the UALib we prove that Lift-alg preserves term identities and is a *homomorphism*, an *algebraic invariant*, and a *sublagebraic invariant*.[41]

**Compatibility of binary relations**

We now define the function compatible so that, if **A** is an algebra and $R$ a binary relation, then compatible **A** $R$ will denote the assertion that $R$ is compatible with all basic operations of **A**. Using the relation $|$: (§3.1) this implication is expressed as $(f\ \hat{}\ $**A**$)\ |:\ R$, yielding a compact representation of compatibility of algebraic operations and binary relations.

    compatible : $($**A** : Algebra $\mathcal{U}\ S) \to$ Rel $\mid$ **A** $\mid\ \mathcal{W} \to \mathcal{O} \sqcup \mathcal{U} \sqcup \mathcal{V} \sqcup \mathcal{W}\ \cdot$
    compatible **A** $R = \forall\ f \to (f\ \hat{}\ $**A**$)\ |:\ R$

**Compatibility of continuous relations\***[29]

In the Relations.Continuous module we defined a function called cont-compatible-op to represent the assertion that a given continuous relation is compatible with a given operation. With that it is easy to define a function, which we call cont-compatible, representing compatibility of a continuous relation with all operations of an algebra. Similarly, we define the analogous dep-compatible function for the (even more general) type of dependent relations.

cont-compatible : $\{I : \mathcal{V}\ \cdot\}($**A** : Algebra $\mathcal{U}\ S) \to$ ContRel $I\mid $**A**$\mid\ \mathcal{W} \to \mathcal{O} \sqcup \mathcal{U} \sqcup \mathcal{V} \sqcup \mathcal{W}\ \cdot$
cont-compatible **A** $R = \Pi\ f :\mid S \mid ,$ cont-compatible-op $(f\ \hat{}\ $**A**$)\ R$

dep-compatible : $\{I : \mathcal{V}\ \cdot\}(\mathcal{A} : I \to$ Algebra $\mathcal{U}\ S) \to$ DepRel $I\ (\lambda\ i \to \mid \mathcal{A}\ i \mid)\ \mathcal{W} \to \mathcal{O} \sqcup \mathcal{U} \sqcup \mathcal{V} \sqcup \mathcal{W}\ \cdot$
dep-compatible $\mathcal{A}\ R = \Pi\ f :\mid S \mid ,$ dep-compatible-op $(\lambda\ i \to f\ \hat{}\ (\mathcal{A}\ i))\ R$

## 4.3 Products

This section presents the Algebras.Products module of the AgdaUALib, slightly abridged.[42]

    Recall the informal definition of a *product* of a family of $S$-algebras. Given a type $I : \mathcal{I}\ \cdot$ and a family $\mathcal{A} : I \to Algebra\ \mathcal{U}\ S$, the *product* $\bigsqcap \mathcal{A}$ is the algebra whose domain is the Cartesian product $\Pi\ i : I\ ,\mid \mathcal{A}\ i \mid$ of the domains of the algebras in $\mathcal{A}$, and the operation symbols are interpreted point-wise in the following sense: if $f$ is a $J$-ary operation symbol and if $a : \Pi\ i : I\ ,\ J \to \mathcal{A}\ i$ gives, for each $i : I$, a $J$-tuple of elements of $\mathcal{A}\ i$, then we define $(f\ \hat{}\ \bigsqcap \mathcal{A})\ a := (i : I) \to (f\ \hat{}\ \mathcal{A}\ i)(a\ i)$. We now define a type that codifies this informal definition of product algebra.

---

[41] See EquationalLogic.html, Homomorphisms.Basic.html, Isomorphisms.html, and Subalgebras.html, resp.
[42] For unabridged docs (source code) see https://ualib.gitlab.io/Algebras.Products.html (https://gitlab.com/ualib/ualib.gitlab.io/-/blob/master/UALib/Algebras/Products.lagda).

$$\prod : (\mathscr{A} : I \to \mathsf{Algebra}\ \mathcal{U}\ S\ ) \to \mathsf{Algebra}\ (\mathcal{I} \sqcup \mathcal{U})\ S$$

$$\prod \mathscr{A} = (\Pi\ i : I\ ,\ |\ \mathscr{A}\ i\ |)\ ,\qquad\text{-- \textit{domain of the product algebra}}$$
$$\qquad\lambda\ f\ a\ i \to (f\ \hat{}\ \mathscr{A}\ i)\ \lambda\ x \to a\ x\ i\ \text{-- \textit{basic operations of the product algebra}}$$

Before going further, let us agree on another convenient notational convention, which is used in many of the later modules of the UALib. Given a signature $S$ : Signature $\mathcal{O}\ \mathcal{V}$, the type Algebra $\mathcal{U}\ S$ has type $\mathcal{O} \sqcup \mathcal{V} \sqcup \mathcal{U}^+$ ·, and $\mathcal{O} \sqcup \mathcal{V}$ remains fixed since $\mathcal{O}$ and $\mathcal{V}$ always denote the universes of operation and arity types, respectively. Such levels occur so often in the UALib that we define the following shorthand for their universes: ov $\mathcal{U} := \mathcal{O} \sqcup \mathcal{V} \sqcup \mathcal{U}^+$.

**Products of classes of algebras**

An arbitrary class $\mathcal{K}$ of algebras is represented as a predicate over the type Algebra $\mathcal{U}\ S$, for some universe level $\mathcal{U}$ and signature $S$. That is, $\mathcal{K}$ : Pred(Algebra $\mathcal{U}\ S$) $\mathcal{W}$ for some $\mathcal{W}$. Later we will formally state and prove that the product of all subalgebras of algebras in $\mathcal{K}$ belongs to the class SP($\mathcal{K}$) of subalgebras of products of algebras in $\mathcal{K}$. That is, $\prod \mathsf{S}(\mathcal{K}) \in \mathsf{SP}(\mathcal{K})$. This turns out to be a nontrivial exercise.

To begin, we need to define types that represent products over arbitrary (non-indexed) families such as $\mathcal{K}$ or $\mathsf{S}(\mathcal{K})$. Observe that $\Pi\ \mathcal{K}$ is certainly not what we want. For recall that Pred(*Algebra* $\mathcal{U}\ S$) $\mathcal{W}$ is just an alias for the function type Algebra $\mathcal{U}\ S \to \mathcal{W}$ ·, and the semantics of the latter takes $\mathcal{K}\ \mathbf{A}$ to mean that $\mathbf{A}$ belongs to the class $\mathcal{K}$. Therefore, by definition

$$\Pi\ \mathcal{K}\ =\ \Pi\ \mathbf{A} : (\mathsf{Algebra}\ \mathcal{U}\ S)\ ,\ \mathcal{K}\ \mathbf{A}\ =\ \forall\ (\mathbf{A} : \mathsf{Algebra}\ \mathcal{U}\ S) \to \mathbf{A} \in \mathcal{K},$$

which denotes the assertion that every inhabitant of the type Algebra $\mathcal{U}\ S$ belongs to $\mathcal{K}$. . Evidently this is not the product algebra that we seek.

What we need is a type that serves to index the class $\mathcal{K}$, and a function $\mathfrak{A}$ that maps an index to the inhabitant of $\mathcal{K}$ at that index. But $\mathcal{K}$ is a predicate (of type (Algebra $\mathcal{U}\ S) \to \mathcal{W}$ ·) and the type Algebra $\mathcal{U}\ S$ seems rather nebulous in that there is no natural indexing class with which to "enumerate" all inhabitants of Algebra $\mathcal{U}\ S$ that belong to $\mathcal{K}$.[43]

The solution is to essentially take $\mathcal{K}$ itself to be the index type; at least heuristically that is how one can think of the type $\mathfrak{I}$ that we now define.[44]

$$\mathfrak{I} : \mathsf{ov}\ \mathcal{U}\ \cdot$$
$$\mathfrak{I} = \Sigma\ \mathbf{A} : (\mathsf{Algebra}\ \mathcal{U}\ S)\ ,\ (\mathbf{A} \in \mathcal{K})$$

Taking the product over the index type $\mathfrak{I}$ requires a function that maps an index $i : \mathfrak{I}$ to the corresponding algebra. Each $i : \mathfrak{I}$ denotes a pair, $(\mathbf{A}\ ,\ p)$, where $\mathbf{A}$ is an algebra and $p$ is a proof that $\mathbf{A}$ belongs to $\mathcal{K}$, so the function mapping such an index to the corresponding algebra is simply the first projection.

$$\mathfrak{A} : \mathfrak{I} \to \mathsf{Algebra}\ \mathcal{U}\ S$$
$$\mathfrak{A} = \lambda\ (i : \mathfrak{I}) \to |\ i\ |$$

Finally, we represent the product of all members of the class $\mathcal{K}$ by the following type.

$$\mathsf{class\text{-}product} : \mathsf{Algebra}\ (\mathsf{ov}\ \mathcal{U})\ S$$
$$\mathsf{class\text{-}product} = \prod \mathfrak{A}$$

---

[43] If you haven't seen this before, give it some thought and see if the correct type comes to you organically.

[44] **Unicode Hints**. Some of our types are denoted with Gothic ("mathfrak") symbols. To produce them in agda2-mode, type \Mf followed by a letter. For example, \MfI ↝ $\mathfrak{I}$.

Observe that the application of 𝔄 to the pair (**A** , *p*) (the result of which is simply the algebra **A**) may be viewed as the projection out of the product ∏ 𝔄 and onto the "(**A**, *p*)-th component" of that product.

## 4.4 Congruences

This section presents the Algebras.Congruences module of the AgdaUALib, slightly abridged.[45]

Recall that a *congruence relation* of an algebra **A** is defined to be an equivalence relation that is compatible with the basic operations of **A**. This concept can be represented in a number of alternative but equivalent ways. Formally, we define a record type (IsCongruence) to represent the *property of being a congruence*, and we define a Sigma type (Con) to represent the *type of congruences* of a given algebra.

```
record IsCongruence (A : Algebra 𝒰 S)(θ : Rel ∣ A ∣ 𝒲) : ov 𝒲 ⊔ 𝒰 ˙ where
  constructor mkcon
  field is-equivalence : IsEquivalence θ
        is-compatible : compatible A θ

Con : (A : Algebra 𝒰 S) → 𝒰 ⊔ ov 𝒲 ˙
Con A = Σ θ : ( Rel ∣ A ∣ 𝒲 ) , IsCongruence A θ
```

Each of these types captures what it means to be a congruence and they are equivalent in the sense that each implies the other. One implication is the "uncurry" operation and the other is the second projection.

```
IsCongruence→Con : {A : Algebra 𝒰 S}(θ : Rel ∣ A ∣ 𝒲) → IsCongruence A θ → Con A
IsCongruence→Con θ p = θ , p

Con→IsCongruence : {A : Algebra 𝒰 S} → ((θ , _) : Con A) → IsCongruence A θ
Con→IsCongruence θ = ∥ θ ∥
```

Above we defined the *zero relation* **0** (3.3) and we now build the *trivial congruence*, which has **0** as its underlying relation. Observe that **0** is equivalent to the identity relation ≡ (§2.2) and these are obviously both equivalence relations. We already proved this of ≡, so we can simply apply the corresponding proofs. The fact that **0** is compatible with all operations of all algebras is equally clear.

```
0-IsEquivalence : {A : 𝒰 ˙} → IsEquivalence {A = A} 0
0-IsEquivalence = record {rfl = refl; sym = ≡-sym; trans = ≡-trans}

0-compatible-op : funext 𝒱 𝒰 → {A : Algebra 𝒰 S} (f : ∣ S ∣) → (f ̂ A) ∣: 0
0-compatible-op fe {A} f {i}{j} ptws0 = ap (f ̂ A) (fe ptws0)

0-compatible : funext 𝒱 𝒰 → {A : Algebra 𝒰 S} → compatible A 0
0-compatible fe {A} = λ f args → 0-compatible-op fe {A} f args
```

Finally, we have the ingredients need to construct the zero congruence of any algebra we like.

```
Δ : (A : Algebra 𝒰 S){fe : funext 𝒱 𝒰} → IsCongruence A 0
Δ A {fe} = mkcon 0-IsEquivalence (0-compatible fe)
```

---

[45] For unabridged docs (source code) see https://ualib.gitlab.io/Algebras.Congruences.html (https://gitlab.com/ualib/ualib.gitlab.io/-/blob/master/UALib/Algebras/Congruences.lagda).

$\mathbb{0}$ : (**A** : Algebra 𝒰 $S$){$fe$ : funext 𝒱 𝒰} → Con{𝒰} **A**
$\mathbb{0}$ **A** {$fe$} = IsCongruence→Con **0** (Δ **A** {$fe$})

### Quotient Algebras

In many areas of abstract mathematics the *quotient* of an algebra **A** with respect to a congruence relation $\theta$ of **A** plays an important role. This quotient is typically denoted by **A** ╱ $\theta$ and Agda allows us to define and express quotients using this standard notation.[46]

_╱_ : (**A** : Algebra 𝒰 $S$) → Con{𝒲} **A** → Algebra (𝒰 ⊔ 𝒲 $^+$) $S$

**A** ╱ $\theta$ = (∣ **A** ∣ / ∣ $\theta$ ∣) ,                 – *the domain of the quotient algebra*
      $\lambda$ $f$ $a$ → ⟪ ($f$ ̂ **A**)($\lambda$ $i$ → fst ∥ $a$ $i$ ∥) ⟫ – *the basic operations of the quotient algebra*

**Example**. Denote by $\mathbb{0}$[ **A** ╱ $\theta$ ] the zero relation on the quotient algebra **A** ╱ $\theta$, which is defined as follows.

$\mathbb{0}$[_╱_] : (**A** : Algebra 𝒰 $S$)($\theta$ : Con {𝒲} **A**) → Rel (∣ **A** ∣ / ∣ $\theta$ ∣)(𝒰 ⊔ 𝒲 $^+$)
$\mathbb{0}$[ **A** ╱ $\theta$ ] = $\lambda$ $u$ $v$ → $u$ ≡ $v$

From this we obtain the zero congruence of **A** ╱ $\theta$ by applying the Δ function defined above.

**0**[_╱_] : (**A** : Algebra 𝒰 $S$)($\theta$ : Con{𝒲} **A**){$fe$ : funext 𝒱 (𝒰 ⊔ 𝒲 $^+$)} → Con (**A** ╱ $\theta$)
**0**[ **A** ╱ $\theta$ ] {$fe$} = $\mathbb{0}$[ **A** ╱ $\theta$ ] , Δ (**A** ╱ $\theta$) {$fe$}

Finally, the following *elimination rule* is sometimes useful.

/-≡ : ($\theta$ : Con{𝒲} **A**){$u$ $v$ : ∣ **A** ∣} → ⟪ $u$ ⟫ {∣ $\theta$ ∣} ≡ ⟪ $v$ ⟫ → ∣ $\theta$ ∣ $u$ $v$
/-≡ $\theta$ refl = IsEquivalence.rfl (is-equivalence ∥ $\theta$ ∥)

## 5 Concluding Remarks

We've reached the end of Part 1 of our three-part series describing the AgdaUALib. Part 2 will cover homomorphism, terms, and subalgebras, and Part 3 will cover free algebras, equational classes of algebras (i.e., varieties), and Birkhoff's HSP theorem.

We conclude by noting that one of our goals is to make computer formalization of mathematics more accessible to mathematicians working in universal algebra and model theory. We welcome feedback from the community and are happy to field questions about the UALib, how it is installed, and how it can be used to prove theorems that are not yet part of the library. Merge requests submitted to the UALib's main gitlab repository are especially welcomed. Please visit the repository at `https://gitlab.com/ualib/ualib.gitlab.io/` and help us improve it.

### References

1   Clifford Bergman. *Universal Algebra: fundamentals and selected topics*, volume 301 of *Pure and Applied Mathematics (Boca Raton)*. CRC Press, Boca Raton, FL, 2012.
2   Ana Bove and Peter Dybjer. *Dependent Types at Work*, pages 57–99. Springer Berlin Heidelberg, Berlin, Heidelberg, 2009. `doi:10.1007/978-3-642-03153-3_2`.

---

[46] **Unicode Hints**. Produce the ╱ symbol in agda2-mode by typing `\---` and then `C-f` a number of times.

**3** Guillaume Brunerie. Truncations and truncated higher inductive types, September 2012. URL: https://homotopytypetheory.org/2012/09/16/truncations-and-truncated-higher-inductive-types/.

**4** Venanzio Capretta. Universal algebra in type theory. In *Theorem proving in higher order logics (Nice, 1999)*, volume 1690 of *Lecture Notes in Comput. Sci.*, pages 131–148. Springer, Berlin, 1999. doi:10.1007/3-540-48256-3_10.

**5** Jesper Carlström. A constructive version of Birkhoff's theorem. *Mathematical Logic Quarterly*, 54(1):27–34, 2008. doi:10.1002/malq.200710023.

**6** Alonzo Church. A formulation of the simple theory of types. *The Journal of Symbolic Logic*, 5(2):56–68, 1940. URL: http://www.jstor.org/stable/2266170.

**7** Jesper Cockx. Dependent pattern matching and proof-relevant unification, 2017. URL: https://lirias.kuleuven.be/retrieve/456787.

**8** William DeMeo. The Agda Universal Algebra Library, Part 2: Structure. *CoRR*, abs/2103.09092, 2021. Source code: https://gitlab.com/ualib/ualib.gitlab.io. arXiv:2103.09092.

**9** William DeMeo. The Agda Universal Algebra Library, Part 3: Identity. *CoRR*, 2021. (to appear) Source code: https://gitlab.com/ualib/ualib.gitlab.io.

**10** Martín Hötzel Escardó. Introduction to univalent foundations of mathematics with Agda. *CoRR*, abs/1911.00580, 2019. arXiv:1911.00580.

**11** Emmanuel Gunther, Alejandro Gadea, and Miguel Pagano. Formalization of universal algebra in Agda. *Electronic Notes in Theoretical Computer Science*, 338:147 – 166, 2018. The 12th Workshop on Logical and Semantic Frameworks, with Applications (LSFA 2017). doi:https://doi.org/10.1016/j.entcs.2018.10.010.

**12** Per Martin-Löf. An intuitionistic theory of types: predicative part. In *Logic Colloquium '73 (Bristol, 1973)*, pages 73–118. Studies in Logic and the Foundations of Mathematics, Vol. 80. North-Holland, Amsterdam, 1975.

**13** nLab authors. Constructive Mathematics. http://ncatlab.org/nlab/show/constructive%20mathematics, March 2021. Revision 65.

**14** nLab authors. Predicative Mathematics. http://ncatlab.org/nlab/show/predicative%20mathematics, March 2021. Revision 22.

**15** nLab authors. Propositions as Types. http://ncatlab.org/nlab/show/propositions%20as%20types, March 2021. Revision 40.

**16** Ulf Norell. Dependently typed programming in agda. In *International school on advanced functional programming*, pages 230–266. Springer, 2008.

**17** Ulf Norell. Dependently typed programming in Agda. In *Proceedings of the 6th International Conference on Advanced Functional Programming*, AFP'08, pages 230–266, Berlin, Heidelberg, 2009. Springer-Verlag. URL: http://dl.acm.org/citation.cfm?id=1813347.1813352.

**18** The Univalent Foundations Program. *Homotopy Type Theory: Univalent Foundations of Mathematics*. Lulu and The Univalent Foundations Program, Institute for Advanced Study, 2013. URL: https://homotopytypetheory.org/book.

**19** Bas Spitters and Eelis Van der Weegen. Type classes for mathematics in type theory. *CoRR*, abs/1102.1323, 2011. arXiv:1102.1323.

**20** The Agda Team. Agda Language Reference section on Axiom K, 2021. URL: https://agda.readthedocs.io/en/v2.6.1/language/without-k.html.

**21** The Agda Team. Agda Language Reference section on Safe Agda, 2021. URL: https://agda.readthedocs.io/en/v2.6.1/language/safe-agda.html#safe-agda.

**22** The Agda Team. Agda Tools Documentation section on Pattern matching and equality, 2021. URL: https://agda.readthedocs.io/en/v2.6.1/tools/command-line-options.html#pattern-matching-and-equality.

**23** The Agda Team. The Fin type of the Agda Standard Library, 2021. URL: https://agda.github.io/agda-stdlib/Data.Fin.html.

**24** Philip Wadler, Wen Kokke, and Jeremy G. Siek. *Programming Language Foundations in Agda*. July 2020. URL: http://plfa.inf.ed.ac.uk/20.07/.